\documentclass[12pt]{iopart}
\usepackage{graphicx}
\usepackage{iopams}
\expandafter\let\csname equation*\endcsname\relax
\expandafter\let\csname endequation*\endcsname\relax
\usepackage{amsmath,amsfonts,amsthm,bm,mathtools}

\usepackage{graphicx}
\usepackage[dvipsnames]{xcolor}
\definecolor{linkcolor}{rgb}{0.3,0.3,1.0} 
\usepackage[pdftex,colorlinks=true, linkcolor= linkcolor, citecolor= linkcolor, urlcolor= linkcolor, hyperindex=true,hyperfigures=true]{hyperref} 

\usepackage{mathpazo,bigints}

\newcommand{\EV}{{\mathcal{S}}}

\newcommand{\Mean}[1]{{\left< {#1} \right>}}
\newcommand{\mean}[1]{{\langle {#1} \rangle}}

\newcommand{\sym}{{^{i	\rightleftarrows j}}}

\newcommand{\MPIPKS}{Max Planck Institute for the Physics of Complex Systems, N{\"o}thnitzer Stra{\ss}e 38, 01187, Dresden, Germany} 
\newcommand{\UNIPD}{
Department of Physics and Astronomy, University of Padova, Via Marzolo 8, I-35131 Padova, Italy}
\newcommand{\INFN}{INFN, Sezione di Padova, Via Marzolo 8, I-35131 Padova, Italy}
\newcommand{\SBL}{Small Biosystems Lab, Condensed Matter Physics Department, Universitat de Barcelona, C/ Marti i Franques 1, 08028 Barcelona, Spain}
\newcommand{\INN}{Institut de Nanoci\`encia i Nanotecnologia (IN2UB), Universitat de Barcelona, 08028 Barcelona, Spain}

\begin{document}

\title{Variance sum rule: proofs and solvable models} 

\author{Ivan Di Terlizzi,$^{1,2,*}$ Marco Baiesi,$^{2,3}$ Felix Ritort$^{4,5}$}
\vspace{3 mm}
\address{$^1$ \MPIPKS}
\address{$^2$ \UNIPD}
\address{$^3$ \INFN}
\address{$^4$ \SBL} 
\address{$^5$ \INN}
\ead{$^*$ ivandt@pks.mpg.de}

\begin{abstract}
We derive, in more general conditions, a recently introduced variance sum rule (VSR) [I. Di Terlizzi et al., 2024 Science {\bf 383} 971] involving variances of displacement and force impulse for overdamped Langevin systems in a nonequilibrium steady state (NESS). This formula allows visualising the effect of nonequilibrium as a deviation of the sum of variances from normal diffusion $2Dt$, with $D$ the diffusion constant and $t$ the time. From the VSR, we also derive formulas for the entropy production rate $\sigma$ that, differently from previous results, involve second-order time derivatives of position correlation functions. This novel feature gives a criterion for discriminating strong nonequilibrium regimes without measuring forces. We then apply and discuss our results to three analytically solved models: a stochastic switching trap, a Brownian vortex, and a Brownian gyrator. Finally, we compare the advantages and limitations of known and novel formulas for $\sigma$ in an overdamped NESS.
\end{abstract}

\section{Introduction}

A non-zero entropy production rate is a key feature of nonequilibrium~\cite{maes03_1,seifert2012stochastic,peliti2021stochastic,shiraishi2023book} that characterizes many physical systems and natural phenomena. For example, it determines the efficiency of energy transduction~\cite{verley2014unlikely,martinez2016brownian,manikandan2019efficiency,landi2021irreversible} and the breakdown of detailed balance in cells \cite{martin2001comparison,battle2016broken,gladrow2016broken,turlier2016equilibrium,lynn2021broken}. For this reason, the study of entropy production is an active and prolific field in stochastic thermodynamics \cite{pigolotti2017generic,gnesotto2018broken,busiello20entropyunidirectional,holubec2020active,busiello2021dissipation,OurGLE,nakazato2021geometrical,freitas2022reliability,venturelli2023stochastic,baiesi2023effective,manzano2023thermodynamics,singh2023inferring} and active matter \cite{pietzonka2017entropy,nardini2017entropy,dabelow2019irreversibility,markovich2021thermodynamics,grandpre2021entropy,fodor2022irreversibility,ro2022model}.
Thanks to technological advancements, it is possible to access microscopic systems where fluctuations are relevant~\cite{ritort2008nonequilibrium,ciliberto2017experiments,seifert2019stochastic,fodor2022irreversibility}, but entropy production remains difficult to measure. To this end, several studies proposed approaches for measuring it in stochastic dynamics. Any trajectory in a diffusive system displays a fluctuating entropy production. According to stochastic energetics~\cite{sek10,sekimoto1998langevin}, such entropy production integrates microscopic forces over displacements, corresponding to the heat delivered to the reservoirs, and it is divided by temperature, as in macroscopic thermodynamics. Other approaches modify this formula by considering gradients of forces~\cite{sek10}. Furthermore, the deviation from the equilibrium fluctuation-dissipation theorem, as in the Harada-Sasa relation~\cite{har05,har06}, also determines entropy production.
Bounds for the entropy production may depend on time irreversibility \cite{andrieux2007entropy,roldan2010estimating,roldan2012entropy,polettini2017effective,manikandan2018exact,martinez2019inferring,li2019quantifying,roldan2021quantifying}.
Other lower bounds derive from thermodynamic uncertainty relations~\cite{bar15,gin16,pietzonka2018universal,macieszczak2018unified,dit19,dechant2020fluctuation,horowitz2020thermodynamic,falasco2020unifying,Di_Terlizzi_Mem_TUR,fu2022thermodynamic,van2022unified} with multiple applications~\cite{hwang2018energetic,song2020thermodynamic,koyuk2020thermodynamic,hartich2021thermodynamic,pal2021thermodynamic,dechant2021improving,pietzonka2023thermodynamic}, optimal bounds \cite{busiello2019hyperaccurate,van2020entropy,van2023thermodynamic} or waiting-time distributions \cite{skinner2021estimating,van2022thermodynamic,ghosal2023entropy,nitzan2023universal}. Coarse-graining ~\cite{esposito2015stochastic,busiello2019entropycoarse,busiello2019entropyCoarseFP,manikandan2020inferring,teza2020exact,ehrich2021tightest,bilotto2021excess,dieball2022mathematical} and partial measurements \cite{shiraishi2015fluctuation,bisker2017hierarchical,polettini2017effective,skinner2021improved,diter24} also influence dissipation estimates. Other studies have addressed colloidal particle models and systems \cite{argun2017experimental,chaki2019effects}

This work derives a multi-dimensional version of a recently introduced variance sum rule (VSR)~\cite{VSR} for measuring the entropy production rate, hereafter referred to as $\sigma$. The VSR gives an exact estimator for $\sigma$ rather than a lower bound. In Ref.~\cite{VSR}, we have shown the performance of the VSR in measuring entropy production in colloidal systems and red blood cells. The VSR sets a methodology to quantify the thermodynamics of life \cite{edgar_perspective}. A model-dependent, reduced-VSR, for example, provides the experimental estimate of the dissipation of active and passivated red blood cells even if partial information on the system is available.  Here, instead, we discuss in more detail the full VSR, in which all relevant degrees of freedom are accessible by measurements.

The VSR (see \eqref{eq:VSR} below) generalises the simple law of free diffusion and visualises the nonequilibrium effects as a deviation from a linear scaling in time. 
Equation \eqref{eq:sigma_VSR} below, its particular case \eqref{eq:sigma2}, and \eqref{eq:sigma3}, are novel formulas for computing the average entropy production rate. Their innovation concerns using a combination of variances of the system's mean square displacement and its instantaneous forces. 
The starting point of our derivation is the standard formula for entropy production as derived in stochastic energetics~\cite{sek10}. An intermediate step \eqref{eq:EP_sekimoto_final} is also a novel formula highlighting how entropy production is related to the time asymmetry of position-force correlation functions.

In the following, we illustrate the main results (section~\ref{sec:VSR}) and present the proof (section~\ref{sec:VSRDer}) of the VSR and its entropy production rate for multi-dimensional overdamped stochastic diffusion dynamics. We consider particle systems with a non-diagonal mobility tensor in contact with baths at different temperatures.
We then illustrate them by solving analytically three models in a steady state driven out of equilibrium by different mechanisms: (i) a particle in a stochastic switching trap, or equivalently, a trapped active particle (section~\ref{sec:SST}), (ii)
a Brownian vortex model with a nonconservative mechanical force (section~\ref{sec:BV}), and (iii) the Brownian gyrator driven by a temperature gradient (section~\ref{sec:Br_gy}). Finally, the discussion section compares previous and new formulas for computing entropy production rates in overdamped systems.

\section{Variance sum rule}\label{sec:VSR}

The VSR holds for $d$-dimensional Markovian systems performing stochastic overdamped diffusion in a nonequilibrium steady state (NESS), for which the probability density function $p(\pmb{x})$ does not change in time. Their coordinates $\pmb{x}=(x^i)$ (with $1 \le i\le d$) evolve with the overdamped Langevin equation
\begin{equation}
\label{eq:LE}
\begin{split}
     \dot{\pmb{x}}_t &
     =  \pmb{\mu} \, \pmb{F}_{t} + \sqrt{2\,\pmb{D}} \cdot \pmb{\xi}_t
\end{split}
\end{equation}
with constant symmetric diffusion matrix $\pmb{D}$ and Gaussian white noise with moments $\langle \xi^{i}_{t'} \rangle = 0$ and $\langle \xi^{i}_{t'} \xi^{j}_{t''} \rangle = \delta_{ij}\, \delta(t'-t'')$. 
The symmetric mobility matrix $\pmb{\mu}$ multiplies the forces $\pmb{F}_t \equiv \pmb{F}(\pmb{x}_{t},t) $, which can have an explicit time dependence on an external protocol $\pmb{\lambda}_t$, provided that it establishes a steady state.
We also assume the second fluctuation-dissipation theorem $\pmb{D} = k_B \pmb{T} \pmb{\mu}$, with diagonal temperature matrix $\pmb{T}$. The Langevin model as described in \eqref{eq:LE} is complementary to other approaches based, such as the Caldeira-Leggett model, where the thermal bath emerges as a collective system of harmonic oscillators interacting with the Brownian particle \cite{cui2018generalized}.

\begin{figure}[t!] 
 	\centering
	\includegraphics[width=0.9
\textwidth]{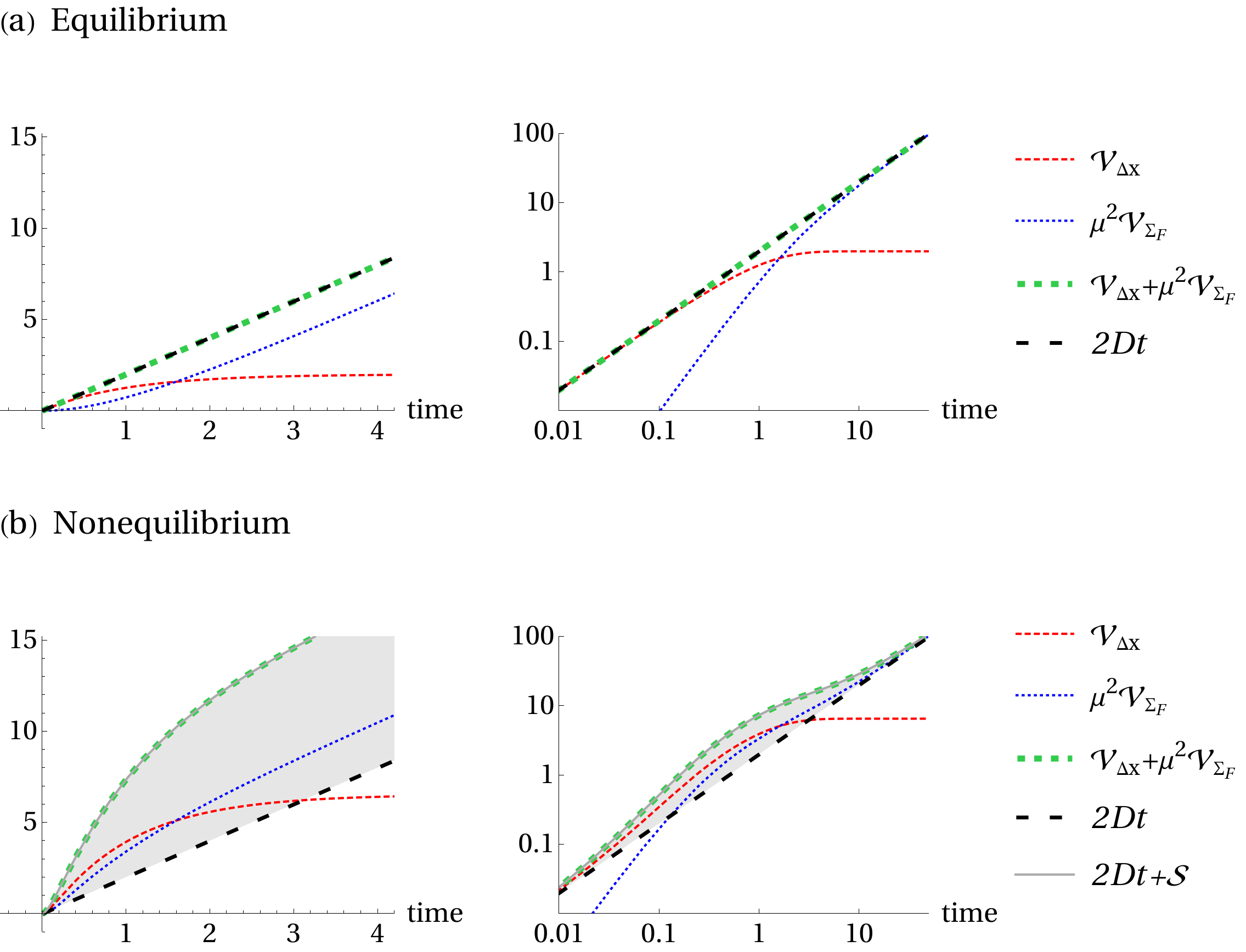}
	\caption{Schematic example, for a one-dimensional system, of the VSR's components (see the legends) for (a) a system in equilibrium and (b) out of equilibrium. We show both linear (left panels) and log-log plots (right panels). The grey-shaded area represents the excess variance.}
	\label{fig:ex}
\end{figure}

In this work, two key dynamical quantities are the displacement $\Delta x_t^i = x_t^i-x_0^i$ and the time-cumulative force or impulse
$\Sigma_F^i(t) = \int_0^t \mathrm{d}t' F^i_{t'}$.
In particular, we focus on extracting information from (co)variances of these observables, which take the form ${\cal V}^{ij}_{A}(t)=\mean{A^i_t A^j_{t}}-\mean{A^i_t}\mean{A^j_t}$, with $\mean{(\cdot)}$ the dynamical average respect to the NESS. One may express these variances in terms of connected correlation functions between variables $A^i_{t'}\equiv A^i(x_{t'})$ and $B^j_{t''}\equiv B^j(x_{t''})$, which, for $t=t'-t''$, are homogeneous in time and are defined as
 \begin{equation}\label{CorrF}
     C_{\!AB}^{ij}(t) = C_{\!AB}^{ij}(t',t'') = \mean{A^i_{t'}B^j_{t''}} - \mean{A^i_{t'}}\mean{B^j_{t''}} \, .
 \end{equation}
In these conditions, and using Einstein's notation for sums over repeated indices, the $d$-dimensional VSR takes the form
\begin{equation}
\label{eq:VSR}
{\cal V}^{ij}_{\Delta x}(t) +\mu_{il} ~\! \mu_{jk} {\cal V}^{lk}_{\Sigma_ F}(t) = 2 ~\! D_{ij} ~\! t + \EV^{ij}(t)
\end{equation}
where 
\begin{align}
    \mathcal{V}^{ij}_{\Delta x}(t) 
    &= \left< {\Delta x}^i_{t}{\Delta x}^j_{t}\right>
    - \left< {\Delta x}^i_{t}\right>\left< {\Delta x}^j_{t}\right> \, ,
    \\
    \mathcal{V}^{ij}_{\Sigma_F} (t)
    &= \int_{0}^t\mathrm{d}t'\int_{0}^t\mathrm{d}t''
    C_{FF}^{ij}(t',t'')
    \, .
\end{align}
We define the {\em excess (co)variance},
\begin{equation}
\label{eq:S} 
\EV^{ij}(t) = 2 \int_{0}^{t}\mathrm{d}t'\, \left[\mu_{jk}\left(C^{ik}_{xF}(t')-C^{ki}_{Fx}(t') \right)\right]\sym \, .
\end{equation}
where we indicate the symmetric part of a matrix $\pmb{C}$ as 
\begin{equation}
    [C^{ij}]\sym \equiv \frac{C^{ij}+C^{ji}} 2
\end{equation}
Clearly, $\pmb{\EV}(t)$ vanishes in equilibrium, where correlation functions are time-symmetric.
However, out of equilibrium $\pmb{\EV}(t) \ne \pmb{0}$.
Figure~\ref{fig:ex} shows a sketch of the VSR in equilibrium and out of equilibrium for a one-dimensional system. The main difference between these cases is the appearance of ${\EV}(t) \ne {0}$ out of equilibrium. Notice the different information conveyed by the linear plots (on the left) and the log-log plots (right).

In the next section, we will prove that the curvature of $\pmb{\EV}(t)$  for $t\to 0$ is related to the entropy production rate $\sigma$ which, in inverse of time units, reads
\begin{equation}
\label{eq:sigmaS}
\sigma = v^{i}\, D^{-1}_{ij}\,v^{j}+
\left.{D^{-1}_{ij}\partial^2_t\EV}^{ij}(t)\right|_{t=0} \, ,
\end{equation}
where $\pmb{v} = \mean{\dot{\pmb{x}}}$ is the stationary mean velocity, and $D^{-1}_{ij} = (D^{-1})_{ij}$.
In terms of variances, we rewrite \eqref{eq:sigmaS} as
\begin{equation}
\label{eq:sigma_VSR}
\sigma = v^{i}\, D^{-1}_{ij}\,v^{j}+\left.\frac{1}{4}{D^{-1}_{ij}\partial^2_t\cal V}^{ij}_{\Delta x}(t)\right|_{t=0} +\frac{1}{2}\mathcal{M}_{ij} {\cal V}^{ij}_{F}\, ,
\end{equation}
where ${\cal V}^{ij}_{F} = \left\langle F^i F^j\right\rangle - \left\langle F^i \right\rangle\left\langle F^j\right\rangle$ is the force variance and we define the matrix $\pmb{\mathcal{M}}$ with components
\begin{equation}
     \mathcal{M}_{lk}=D^{\text{-}1}_{ij} \, \mu_{il} \, \mu_{jk} = k_B^{-1}T^{-1}_{ir}\,\mu^{-1}_{rj} \, \mu_{jk}  \, \mu_{il} = k_B^{-1}T^{-1}_{ki} \, \mu_{il}\, ,
\end{equation}
(see further assumptions in the next session).

The solved models in the following sections have diagonal mobility and temperature matrices. Hence, entropy production collects diagonal contributions from \eqref{eq:sigma_VSR}.
By denoting the diagonal components with $\mu_i\equiv \mu_{ii}$ and $T_i\equiv T_{ii}$, the entropy production rate \eqref{eq:sigma_VSR} is rewritten as
\begin{equation}\label{eq:sigma2}
    \sigma 
    ~= 
    \sum_{i=1}^d \sigma_i
    ~=
    \sum_{i=1}^d
    \frac 1 {k_B T_i}
    \left[
    \frac{1}{\mu_i} ({v}^i)^2 
    +
    \left. 
    \frac{1}{4 \mu_i}  \partial_{t}^{2}
    \mathcal{V}^{i}_{\Delta x}(t)
    \right|_{t=0}
    + 
    \frac{\mu_i}{2}
    \mathcal{V}^{i}_{F} 
    \right]
    \,.
\end{equation}
Here, for the variances of the displacement $\Delta x^i$ and of the instantaneous force $F^i$, we have introduced the notation
\begin{align} \label{eq:DX---}
    \mathcal{V}^{i}_{\Delta x}(t) 
    &= \left< ({\Delta x}^i_{t})^2\right>
    - \left< {\Delta x}^i_{t}\right>^2\,,
    \\
    \mathcal{V}^{i}_{F} 
    &= \langle (F^i)^2\rangle
    - \langle F^i\rangle^2\,.
\end{align}
Interestingly, in \eqref{eq:sigma2}, the terms involving second derivatives are the only ones that can be negative. In particular, they compensate the positive force variances at equilibrium to yield $\sigma=0$. As a consequence, we have the inequality
\begin{equation}\label{eq:sigma2inf}
    \sigma \ge \max(\sigma_c, 0)
    \qquad{\textrm{with}}\quad
    \sigma_c \equiv \sum_{i=1}^d
    \frac 1 {4 k_B T_i \mu_i}
    \left.  \partial_{t}^{2}
    \mathcal{V}^{i}_{\Delta x}(t)
    \right|_{t=0}
\end{equation}
Such inequality suggests the condition
\begin{equation}
\label{eq:sigma2inf2}
\sigma_c > 0
\end{equation}
as a measure of a strong departure from equilibrium. In practice, if relevant degrees of freedom $x^i$ are accessible in a Markovian system, while forces are not measurable, from \eqref{eq:sigma2inf}, one can still deduce that the system is significantly far from equilibrium, even if all mean drift velocities  ${v}^i =0$.

Before starting with the derivations, we recall
that associated to the LE~\eqref{eq:LE} there is the Fokker-Planck equation for the probability density $p_t \equiv p(\pmb{x}_t,t)$,
\begin{equation}\label{FP_1}
    \partial_{t} p_t =- \nabla \cdot \pmb{J}_t = - \nabla \cdot (\pmb{\nu}_t \, p_t) \, ,
\end{equation}
where $\pmb{J}_t \equiv \pmb{J}(x_t,t)$ is the probability flux and $\pmb{\nu}_t \equiv \pmb{\nu}(x_t,t)$ is the local mean velocity $\pmb{\nu}_t$,
\begin{equation}
    \label{eq:nu}
    \pmb{\nu}_t = \pmb{\mu F}_t - \pmb{D} ~ \nabla  \ln  p_t  \, .
\end{equation}
The excess variance can be related also to $\pmb{\nu}$,
\begin{equation}\label{eq:Sv} 
\EV_{ij}(t) = 4 \int_{0}^{t}\mathrm{d}t'\int_{0}^{t'}\mathrm{d}t''\, \left[C^{ij}_{\dot x \nu }(t'') \right]\sym\, 
\end{equation}
as explained in 
\ref{app:Deriv_vel} (using the results in \ref{app:R}, and the proof of null covariance between position and mean local velocity in \ref{app:y}).
In one dimension, a {\em violation factor} related to the excess variance was previously introduced in \cite{spe06}, where it was shown that its magnitude measures the amount of violation of the fluctuation-dissipation theorem. This confirms that $\pmb{\EV}$ relates to the distance from equilibrium and, thus, to entropy production and time reversal symmetry.

\section{Derivations}
\label{sec:VSRDer}

We derive the formulas exposed in the previous section for systems evolving according to the stochastic overdamped dynamics described by \eqref{eq:LE}. Each degree of freedom $i$ is in contact with a heat bath at temperature $T_i$. Hence, the temperature matrix is diagonal $(\pmb{T})_{ij} = T_i \delta_{ij}$.  
We restrict our domain to two cases in which the diffusion matrix $\pmb{D}$ can be written as a product of $\pmb{T}$ and a mobility matrix $\pmb{\mu}$:
\begin{itemize}
    \item[i)] The system is in contact with only one heat bath at temperature $T$, i.e. $\pmb{T} = T\,\mathbb{1} $. As a consequence,  $\pmb{D} = k_B T \pmb{\mu}$, with $\pmb{\mu}$ symmetric;
    \item[ii)] $T_i$'s are different and $(\pmb{\mu})_{ij} = \mu_i  \, \delta_{ij}$  is diagonal, implying that $(\pmb{D})_{ij} = k_B T_i \, \mu_i \, \delta_{ij}$.
\end{itemize}
To generate a NESS,  the force could be a time-independent, nonconservative function of the position, $\pmb{F}_{t}=\pmb{F}(x_t)$. In other cases, it could take the form $\pmb{F}_{t}= -\nabla U(\pmb{x}_t-\pmb{v} t)$ if $U(x)$ is a potential energy and $\pmb{v}$ a constant velocity. For example, imagine a harmonic optical trap whose centre moves as $\pmb{\lambda_t} = \pmb{v} t$. 
Another case leading to a steady state is when $\pmb{\lambda}_t$ also performs an autonomous stochastic motion (i.e., not depending on $\pmb{x}_t$), as in the model of section~\ref{sec:SST}. 

We denote time derivatives of the correlation function \eqref{CorrF} as
 $\overset{\,\bm.}{C}\,{\vphantom{C}}_{\!AB}^{ij}(t)  \equiv \partial_t C_{AB}^{ij}(t)$. Their limit
$\overset{\,\bm .}{C}\,{\vphantom{C}}_{\!AB}^{ij}(0^+) = \lim_{t\to 0^+} \overset{\,\bm .}{C}\,{\vphantom{C}}_{\!AB}^{ij}(t) $ is always taken for positive $t$ approaching zero. Also, in case $i=j$, the notation is simplified to $ C_{\!AB}^{i}= C_{\!AB}^{ii}$.

\subsection{VSR}

We start from the time integral of equation \eqref{eq:LE} with rearranged terms,
\begin{equation}
    \Delta\pmb{\mathcal{R}}_{t} \equiv\pmb{x}_t-\pmb{x}_0 -  \int_{0}^{t}\mathrm{d}t'  \pmb{\mu}~\!\pmb{F}_{t'} = \sqrt{2~\!\pmb{D}}\int_{0}^{t} \mathrm{d}t'   \pmb{\xi}_{t'}  \,,
\end{equation}
where we defined $\Delta\pmb{\mathcal{R}}_{t}$, which is useful for calculations. Of course $\mean{\Delta\pmb{\mathcal{R}}_{t}}=0$, while components of its covariance are (with $[C^{ij}]\sym = (C^{ij}+C^{ji})/2$)
\begin{equation}\label{Sum_rule_step1}
{\cal V}^{ij}_{\Delta \mathcal{R}}     =
    {\cal V}^{ij}_{\Delta x}(t) +\mu_{il} ~\! \mu_{jk}{\cal V}^{lk}_{\Sigma_ F}(t)   - 2 \int_{0}^{t}\mathrm{d}t' \left[\mu_{jk}\, \mathrm{Cov} \left( x^{i}_{t}-x^{i}_{0} ~\!,~\! F^{k}_{t'} \right)\right]\sym  =  2 ~\! D_{ij} ~\! t \, ,
\end{equation}
with (co)variances ${\cal V}^{ij}_{A}(t)=\mean{A^i_t A^j_{t}}-\mean{A^i_t}\mean{A^j_t}$ defined in the previous section and where, for the last term, we used the properties of the Gaussian white noise. We also continue to use Einstein's notation for repeated indexes. To complete the proof, we exploit the time homogeneity of the correlation functions to rewrite equation \eqref{Sum_rule_step1} in a form equivalent to \eqref{eq:VSR},
\begin{equation}\label{Sum_rule_step2}
{\cal V}^{ij}_{\Delta x}(t) +\mu_{il} ~\! \mu_{jk} {\cal V}^{lk}_{\Sigma_ F}(t) = 2 ~\! D_{ij} ~\! t +2 \int_{0}^{t}\mathrm{d}t'\left[\mu_{jk}\left(C^{ik}_{xF}(t')-C^{ki}_{Fx}(t') \right)\right]\sym \,.
\end{equation}

\subsection{Entropy production}

The mean entropy production rate $\sigma$ is constant in a NESS, hence the entropy $\sigma\,dt$ produced on average in an infinitesimal time interval $dt$ does not depend on the observation time $t'$. According to stochastic energetics \cite{sek10}, for a system described by \eqref{eq:LE}, in $k_B$ units for $\sigma$,
\begin{equation}
\sigma\,dt = \sum_i \frac{1}{k_B T_i}\mean{F^i_{t'}\circ {dx}^{i}_{t'}}
\label{eq:EP_sekimoto}   
\end{equation}
where $T_i$ is the temperature associated to the $i^{th}$ degree of freedom, and $F^i_{t'}\circ {dx}^{i}_{t'}$ is the Stratonovich product between the force in the interval $[t',t'+dt]$ and the displacement ${dx}^{i}_{t'}$ during that time step. Focusing on the average product of force and infinitesimal displacement, corresponding to the average infinitesimal amount of heat $\mean{dQ^i} = \mean{F^i_{t'}\circ d{x}^{i}_{t'}} $ going to reservoir $i$ in a time $dt$, we can write
\begin{equation}\label{eq:sek_der}
\begin{split}
    \mean{F^i_{t'}\circ d{x}^{i}_{t'}} 
    &= \frac 1 2 \Mean{\left(F^i_{t'+dt}+F^i_{t'}\right)\left(x^i_{t'+dt}-x^i_{t'}\right)}\\
    &= \frac 1 2 \Mean{\left(F^i_{t'+dt}+F^i_{t'}\right)\left(y^i_{t'+dt}-y^i_{t'} + v^i dt\right)}\\
    &= \frac 1 2 \left[ C_{yF}^{i}(dt)-C_{Fy}^{i}(dt)\right] +  v^{i}\, \mu^{-1}_{ij}\,
    v^j \,dt
\end{split}
\end{equation}
where $y^i_t = x^i_t-v^i t$ is the position in the frame moving at constant speed $v^i$, where correlation functions are indeed homogeneous in time.
Also, in \eqref{eq:sek_der} we used that equation \eqref{eq:LE} implies $ \mu_{ji}\mean{F^i_{t'}}=v^j$,
and $\mu^{-1}_{ij}=(\pmb{\mu}^{-1})_{ij}$.
By performing a Taylor expansion up to order $dt$, one gets
\begin{equation}\label{eq:sek_der_3}
\begin{split}
    \mean{F^i_{t'}\circ d{x}^{i}_{t'}} 
    & = \frac 1 2 \left[ \overset{\,\bm .}{C}\,{\vphantom{C}}_{\!yF}^{i}(0^+)-\overset{\,\bm .}{C}\,{\vphantom{C}}_{\!Fy}^{i}(0^+)\right] dt +  
    v^{i}\, \mu^{-1}_{ij}\,v^j \,dt \\
    & = \frac 1 2 \left[ \overset{\,\bm .}{C}\,{\vphantom{C}}_{\!xF}^{i}(0^+)-\overset{\,\bm .}{C}\,{\vphantom{C}}_{\!Fx}^{i}(0^+)\right] dt +  
    v^{i}\, \mu^{-1}_{ij}\,v^j \,dt \, .
\end{split}
\end{equation}
Here, we use that correlation functions involving $y^i_t$ are equal to those involving $x^i_t$ up to a constant scaling linearly with time.
With \eqref{eq:sek_der_3} we rewrite \eqref{eq:EP_sekimoto} as
\begin{equation}
\sigma = v^{i}\, D^{-1}_{ij}\,
    v^j + \sum_i \frac{1}{2k_B T_i}\left[\overset{\,\bm .}{C}\,{\vphantom{C}}_{\!xF}^{i}(0^+)-\overset{\,\bm .}{C}\,{\vphantom{C}}_{\!Fx}^{i}(0^+)\right] \,
\label{eq:EP_sekimoto_final}    \end{equation}
where $D^{-1}_{ij}=(\pmb{D}^{-1})_{ij}$ and $\pmb{D}$ satisfies one of the two conditions i) and ii) listed at the beginning of this section. To our knowledge, \eqref{eq:EP_sekimoto_final} is a novel result, connecting $\sigma$ in \eqref{eq:EP_sekimoto} with the time asymmetry of the position-force correlation functions. This formula is the intermediate step in the derivation of our main results, continued hereafter.

\subsection{Entropy production from the VSR}\label{SS2:EPfromVSR}

 Consider now the multidimensional VSR \eqref{Sum_rule_step2}, multiply both sides by $D^{-1}_{ij}$ and sum over $i,j$, hence obtaining, in Einstein notation,
 \begin{equation}\label{Sum_rule_entropy}
{D^{-1}_{ij}\cal V}^{ij}_{\Delta x}(t) +\mathcal{M}_{ij} {\cal V}^{ij}_{\Sigma_ F}(t) = 2  d  t +2 D^{-1}_{ij} \int_{0}^{t}\mathrm{d}t'\, \left[\mu_{jk}\left(C^{ik}_{xF}(t')-C^{ki}_{Fx}(t') \right)\right]\sym \, ,
\end{equation}
where $d$ is the total number of degrees of freedom and 
\begin{equation}
    \mathcal{M}_{ij}=D^{-1}_{kl}\mu_{li} \, \mu_{kj} \,.
    \label{eq:M}
\end{equation}
In $d=1$, the latter corresponds to $\mathcal{M}=\mu/k_B T$. Because of the hypotheses made on the diffusion matrix, it holds that $k_B \,D^{-1}_{ik}\mu_{kj}= (\pmb{T}^{-1})_{ij}\equiv T^{-1}_{ij}$, implying that the last term in \eqref{Sum_rule_entropy} becomes 
\begin{equation}\label{eq:ent_VSR_1}
    2 D^{-1}_{ij} \int_{0}^{t}\mathrm{d}t'\, \left[\mu_{jk}\left(C^{ik}_{xF}(t')-C^{ki}_{Fx}(t') \right)\right]\sym = 2 k_B^{-1} T^{-1}_{ij}\int_{0}^{t}\mathrm{d}t'\, \left(C^{ij}_{xF}(t')-C^{ji}_{Fx}(t') \right)\, .
\end{equation}
 Since the latter is diagonal, one can rewrite equation \eqref{eq:ent_VSR_1} as
\begin{equation}\label{eq:ent_VSR_2}
    2 k_B^{-1} T^{-1}_{ij}\int_{0}^{t}\mathrm{d}t'\, \left(C^{ij}_{xF}(t')-C^{ji}_{Fx}(t') \right) = \sum_i \frac{2}{k_B T_i}\int_{0}^{t}\mathrm{d}t'\, \left(C^{i}_{xF}(t')-C^{i}_{Fx}(t') \right) \, .
\end{equation}
Finally, by taking the second derivative evaluated at $t=0$ of \eqref{Sum_rule_entropy} and using \eqref{eq:EP_sekimoto_final}, \eqref{eq:M}, and \eqref{eq:ent_VSR_2}, one immediately sees that 
\begin{equation}
    \sigma = v^{i}\, D^{-1}_{ij}\,v^{j}+\frac{1}{4}{D^{-1}_{ij}\partial^2_t\cal V}^{ij}_{\Delta x}(t)|_{t=0} +\frac{1}{2}\mathcal{M}_{ij} {\cal V}^{ij}_{F}\; ,
\end{equation}
because $\mathcal{M}_{ij}\partial^2_t {\cal V}^{ij}_{\Sigma_ F}(t)|_{t=0}= 2\mathcal{M}_{ij} {\cal V}^{ij}_{F}$. Hence, the latter equation corresponds to \eqref{eq:sigma_VSR}.

\section{Stochastic switching trap}
\label{sec:SST}

\begin{figure}[t!] 
 	\centering
	\includegraphics[width=0.98\textwidth]{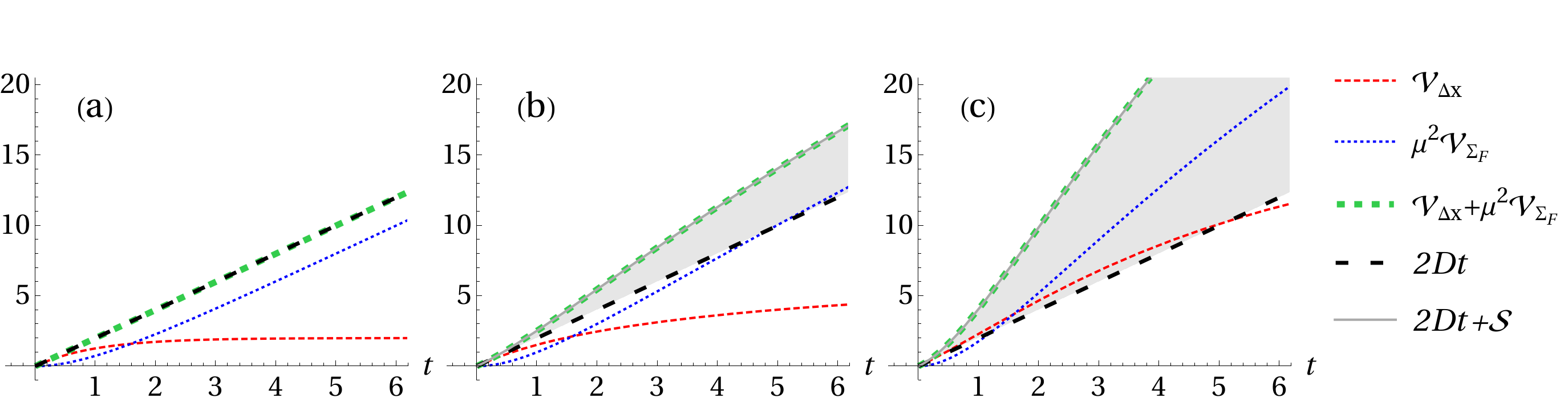}
	\caption{For the stochastic switching trap, various quantities in the VSR vs time (see legend), for (a) ${\epsilon}=0$ (equilibrium) (b)  ${\epsilon}=3/2$, and (c)  ${\epsilon}=3$.
	Common parameters are $w=2/5$,  $k_{B}T=\mu=\kappa=1$. The gray area represents a positive $\mathcal{S}(t)$.}
	\label{fig:VSRsst}
\end{figure}

In this Section, we start the illustration of the VSR with analytically solved models.
First, we provide more details on a model presented in~\cite{VSR} where a
Brownian particle with mobility $\mu$, in water at a temperature $T$, is driven by a harmonic trap whose center $\lambda_t$ jumps stochastically between the positions $\{0,\Delta\lambda\}$. The potential energy thus is $U(x_t,\theta_t) = \kappa \left(x_{t}-\Delta\lambda\,\theta_{t}  \right)^{2}/2$, with dichotomous stochastic variable $\theta_t = \{0,1\}$. The trap undergoes a Markovian jumping dynamics with jumping rates $w_0$ for the $0\to 1$ transition and $w_1$ for the reverse one. Each jump of $\theta_t$ changes instantaneously the particle's potential energy by performing a mechanical work that, on average, is positive. This injected mechanical power equals the average heat flux dissipating energy to the bath, which leads to a positive average entropy production rate $\sigma$.
The stationary average of $\theta_{t}$ can be written as $q = \langle \theta_{t} \rangle =w_0/w$, where $w = w_0+w_1$. 
A Langevin equation models the dynamics of the particle's position,
\begin{equation}\label{LE_switch_present}
     \dot{x}_t
      = -\mu \kappa \left( x_t - \Delta\lambda \theta_t\right) + \sqrt{2\,k_{B}T\,\mu}~ \xi_t \,,
\end{equation}
with $\langle \xi_{t'} \rangle = 0$ and $\langle \xi_{t'} \xi_{t''} \rangle = \delta(t'-t'')$.
We will show how to calculate relevant quantities in terms of stationary connected correlation functions $C_{xx}(t)$, $C_{x\theta}(t)$, $C_{\theta x}(t)$ and $C_{\theta\theta}(t)$. To compute these correlations, we turn to a fine time-step description of the dynamics,
\begin{subequations}
\label{LE_switch_disc}
\begin{align}
x_{t+dt} &= x_t -\mu\,\kappa\,x_t\,dt +\mu\,{\kappa}\,\Delta\lambda\,\theta_t\,dt + \sqrt{ 2\,k_{B}T\,\mu} \,d\mathcal{B}^{x}_t
\label{LE_switch_disc_a}
\\
\theta_{t+dt} &= \theta_t +(1-2\,\theta_t)\Theta(w_{\theta_{t}}dt -r)
\label{LE_switch_disc_b}
\end{align}
\end{subequations}
where $r$ is random variable with uniform probability distribution on $[0,1]$ and $\Theta(\cdot)$ is the Heaviside step function. 
By multiplying \eqref{LE_switch_disc_a} respectively by $x_0$ or $\theta_0$, then taking stationary averages, and removing products of $\mean{x_t}= \mean{x_0}= q\,\Delta\lambda$, $\mean{\theta_t}=\mean{\theta_0}=q$, one obtains 
\begin{subequations}
\label{eq:dCss}
\begin{align}
\label{corr_1_ch6}
    \partial_{t} C_{xx}(t) 
    &= - \mu \,\kappa\,C_{xx}(t) + \mu\, {\kappa\,\Delta\lambda} \,C_{\theta x}(t) \,,
    \\
    \label{corr_2_ch6}
    \partial_{t} C_{x \theta}(t) 
    &= - \mu\,\kappa\,C_{x\theta}(t) + \mu\,{\kappa\,\Delta\lambda} \,C_{\theta \theta}(t)
      \,.
\end{align}
In a similar way, by multiplying \eqref{LE_switch_disc_b} by $x_0$ or $\theta_0$, we get to
\begin{align}\label{corr_3_ch6}
    \partial_{t} C_{\theta x}(t) 
    &= - w \,C_{\theta x}(t) \,,
    \\
    \label{corr_4_ch6}
    \partial_{t} C_{\theta \theta}(t) 
    &= -w\,C_{\theta \theta}(t) \,.
\end{align}
\end{subequations}
Similarly, the stationary initial conditions
\begin{equation}
\begin{split}
    C_{xx}(0) &= \frac{k_B \, T}{\kappa}+\frac{\kappa\,{\Delta \lambda}^{2}\mu \,q(1-q)}{(w+\mu\,\kappa)} \\
    C_{x\theta}(0) &= 
\frac{{\kappa\,\Delta\lambda}\,\mu \,q(1-q)}{w+\mu\,\kappa} \,\,= C_{\theta x}(0) \\
    C_{\theta\theta}(0) &= q(1-q) \, .
\end{split}
\end{equation}
emerge from terms of order $dt$ in the self and cross-products of equations \eqref{LE_switch_disc}.
With these initial conditions, we solve  \eqref{eq:dCss} with standard techniques, obtaining
\begin{equation}
    \begin{split}
    C_{xx}(t) = & \left(\frac{k_B \, T}{\kappa}+\frac{{\kappa\,\Delta\lambda}^{2}\,\mu \,q(1-q)}{(w+\mu\,\kappa)}\right)\text{e}^{-\mu\,\kappa\,t} + 
\frac{({\kappa\,\Delta\lambda})^{2}\,\mu^{2} \,q(1-q)}{w^{2}-\mu^{2}\,\kappa^{2}} \left(\text{e}^{-\mu\,\kappa\,t}-\text{e}^{-w\,t} \right) \\[6pt]
    C_{x\theta}(t) =& \frac{{\kappa\,\Delta\lambda}\,\mu \,q(1-q)}{w+\mu\,\kappa} \text{e}^{-\mu\,\kappa\,t} + \frac{{\kappa\,\Delta\lambda}\,\mu \,q(1-q)}{w-\mu\,\kappa} \left(\text{e}^{-\mu\,\kappa\,t}-\text{e}^{-w\,t} \right)  \\[6pt]
    C_{\theta x}(t)  =& \frac{{\kappa\,\Delta\lambda}\,\mu \,q(1-q)}{w+\mu\,\kappa} \text{e}^{-\mu\,\kappa\,t} \\[6pt]
    C_{\theta\theta}(t) =& q(1-q)\text{e}^{-w\,t} \, .
\end{split}
\end{equation}
From these correlation functions, and by defining $\epsilon = \kappa \Delta\lambda\sqrt{q(1-q)}$, one can compute the variance of the relative displacement, that is
\begin{equation}\label{switch_var_pos}
\begin{split}
    {\cal V}_{\Delta x_{t}} 
    =& \langle (x_{t}-x_{0})^2\rangle
    \\[6pt]
    =& 2 (C_{xx}(0)-C_{xx}(t))\\[6pt]
    =& 2\left[\left(\frac{k_B \, T}{\kappa}+\frac{{\epsilon}^{2}\,\mu}{\kappa(w+\mu\,\kappa)}\right)(1-\text{e}^{-\mu\,\kappa\,t}) 
    + \frac{{\epsilon}^{2}\,\mu^{2}}{w^{2}-\mu^{2}\,\kappa^{2}} \left(\text{e}^{-w\,t}-\text{e}^{-\mu\,\kappa\,t} \right)\right]\, .
\end{split}
\end{equation}
Note that ${\epsilon}$ can be interpreted as the strength of an active force~\cite{VSR} and for ${\epsilon}=0$, one recovers the usual formula for the relative displacement of the equilibrium Ornstein–Uhlenbeck process, $C_{xx}(t)=(k_B T/\kappa)e^{-\mu \kappa t}$. The second term we need for the VSR is the variance of the time integral of the force
$F_t=-\kappa \, x_t + {\kappa\,\Delta\lambda}\, \theta_t$. The symmetry of the correlation functions $\langle F_{t'}\,F_{t''} \rangle$ grants that 
\begin{equation}
{\cal V}_{\Sigma_F}(t)=\int_{0}^{t}\mathrm{dt'}\int_{0}^{t}\mathrm{dt''}\,\langle F_{t'}\,F_{t''} \rangle = 2\int_{0}^{t}\mathrm{dt'}\int_{0}^{t'}\mathrm{dt''}\,\langle F_{t''}\,F_{0} \rangle \, .
\end{equation}
This allows computing
\begin{equation}\label{switch_var_sum_forc}
\begin{split}
    {\cal V}_{\Sigma_F}(t)=& 2 \int_{0}^{t}\mathrm{dt'}\int_{0}^{t'}\mathrm{dt''} \, \left( \kappa^{2}C_{xx}(t'') + (\kappa\,\Delta\lambda)^{2}C_{\theta\theta}(t'') - \kappa^2\Delta\lambda (C_{x\theta}(t'') +C_{\theta x}(t'') ) \right)\\[6pt]
    =& \frac{2}{\mu\,\kappa}\left[\frac{k_{B}T}{\mu}\left(\mu \kappa t+ 1- \text{e}^{-\mu\,\kappa\,t} \right)+\frac{\epsilon^{2}}{(w+\mu\,\kappa)}\left(1 - \frac{w\,\text{e}^{-\mu\,\kappa\,t}}{w-\mu\,\kappa}+ \frac{\mu\,\kappa\,\text{e}^{-w\,t}}{w-\mu\,\kappa} \right)\right]
\end{split}
\end{equation}
We may compute $\EV(t)$ with the VSR and equations \eqref{switch_var_pos} and \eqref{switch_var_sum_forc}, 
\begin{equation}\label{viol_fac_switch}
    \begin{split}
     \EV(t) =&    {\cal V}_{\Delta x_{t}}(t)+ \mu^{2}{\cal V}_{\Sigma_F}(t) - 2~\!k_{B}T~\!\mu~\!t\\[6pt]
     =& \frac{4~\!{\epsilon}^{2} }{\kappa(w+\mu~\!\kappa)}\left(1 - \frac{w~\!\text{e}^{-\mu~\!\kappa~\!t}}{w-\mu~\!\kappa}+ \frac{\mu~\!\kappa~\!\text{e}^{-w~\!t}}{w-\mu~\!\kappa} \right)\,.
    \end{split}
\end{equation}
An example of the temporal evolution of all terms of the VSR derived above is shown in Figure \ref{fig:VSRsst}.

\begin{figure}[t!] 
 	\centering
	\includegraphics[width=0.45\textwidth]{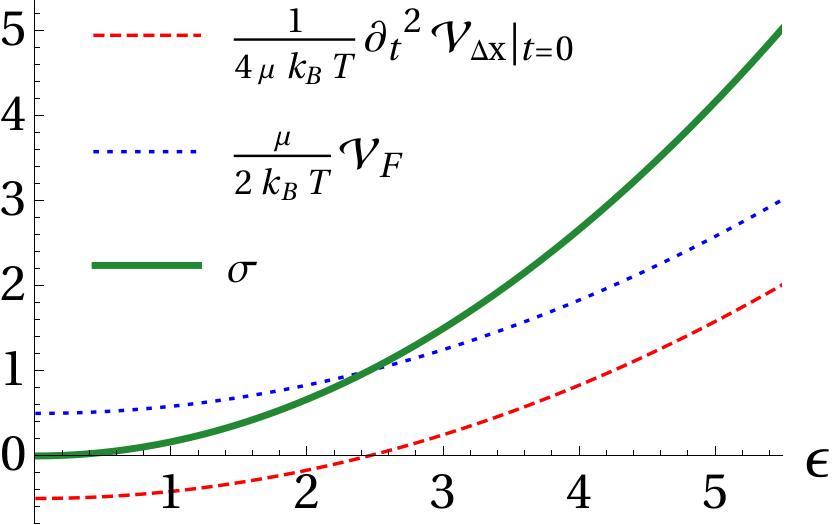}
	\caption{Entropy production for a particle in a stochastic switching trap as a function of the nonequilibrium strength ${\epsilon}=\kappa \Delta\lambda \sqrt{q(1-q)}$. The parameters are the same of figure~\ref{fig:VSRsst}.}
	\label{fig:sig-sst}
\end{figure}

Finally, we can also compute the average entropy production rate in a steady state (with no drift, $v=0$). From \eqref{eq:sigma2} we get,
\begin{align}
    \frac{1}{4 \mu k_B T}  \partial_{t}^2
    \mathcal{V}_{\Delta x}(t)|_{t=0}
    & = 
    -\kappa  \mu+ \frac{\epsilon^2 \, \mu }{2k_BT(1+\kappa  \mu/w)}
    \nonumber\\[5pt]
    \frac{\mu}{2 k_B T}\, \mathcal{V}_F
    &=
    \kappa  \mu + \frac{\epsilon^2 \, \mu }{2k_BT(1+\kappa  \mu/w)} \, ,
\end{align}
hence
\begin{equation}
    \sigma 
    = \frac{1}{4 \mu k_B T}  \partial_{t}^2
    \mathcal{V}_{\Delta x}(t)|_{t=0}+ \frac{\mu}{2 k_B T}\, \mathcal{V}_F =
     \frac{\epsilon^2 \, \mu }{k_BT(1+\kappa  \mu/w)}
\end{equation}
(see also~\cite{garcia2021run}).
As expected, $\sigma$ is zero for ${\epsilon}=0$, namely in equilibrium, where the two terms $\frac{1}{4 \mu k_B T}  \partial_{t}^{2}\mathcal{V}_{\Delta x}(t)|_{t=0} = -\kappa  \mu$ and $\frac{\mu}{2 k_BT}\, \mathcal{V}_F =\kappa  \mu$ cancel each other. As also shown in Figure~\ref{fig:sig-sst}, by going gradually further away from equilibrium, one reaches a value of ${\epsilon}$ where the curvature $\partial_{t}^{2}\mathcal{V}_{\Delta x}(t)|_{t=0}$ changes sign and there starts a regime of positive curvature that is, according to the criterion~\eqref{eq:sigma2inf2}, a clear sign of nonequilibrium conditions.

We can also check that this expression matches $\sigma$ from the stochastic energetics equation \eqref{eq:EP_sekimoto},
\begin{equation}
\begin{split}
     \sigma 
     =& 
      \frac{1}{k_B T}\Big\langle F_t\circ \dot x_{t} \Big\rangle\\[6pt]
     =&\frac{\kappa}{k_B T}\Big\langle -x_{t}\circ \dot x_{t} + \Delta\lambda\,\theta_{t}\circ \dot x_{t} \Big\rangle
     \\[6pt]
     =& 
     \frac{\kappa}{k_B T}\left[- \frac{1}{2} \frac{d}{dt} \langle x_{t}^{2} \rangle +\mu\,\Delta\lambda\,\Big\langle \theta_{t}\circ\Big( -\kappa \, x_{t} + \kappa\,\Delta\lambda \theta_{t}  + \sqrt{2\,k_{B}T/\mu}~ \xi_t \Big)\Big\rangle\right]\\[6pt]
     = & \frac{\mu\,\kappa^2\,\Delta\lambda}{k_B T} \left[\Delta\lambda\,C_{\theta \theta}(0)-C_{\theta x}(0) \right] \\[6pt]
     = & 
     \frac{\epsilon^2\,\mu }{k_BT(1+\kappa  \mu/w)}
\end{split}
\end{equation}
where between the last two lines we used that $\langle x_{t}^{2} \rangle$ is constant and that $\langle \theta_t \circ \xi_t \rangle=0$.

\section{Brownian vortex}
\label{sec:BV}

\begin{figure}[t!] 
 	\centering
	\includegraphics[width=0.5\textwidth]{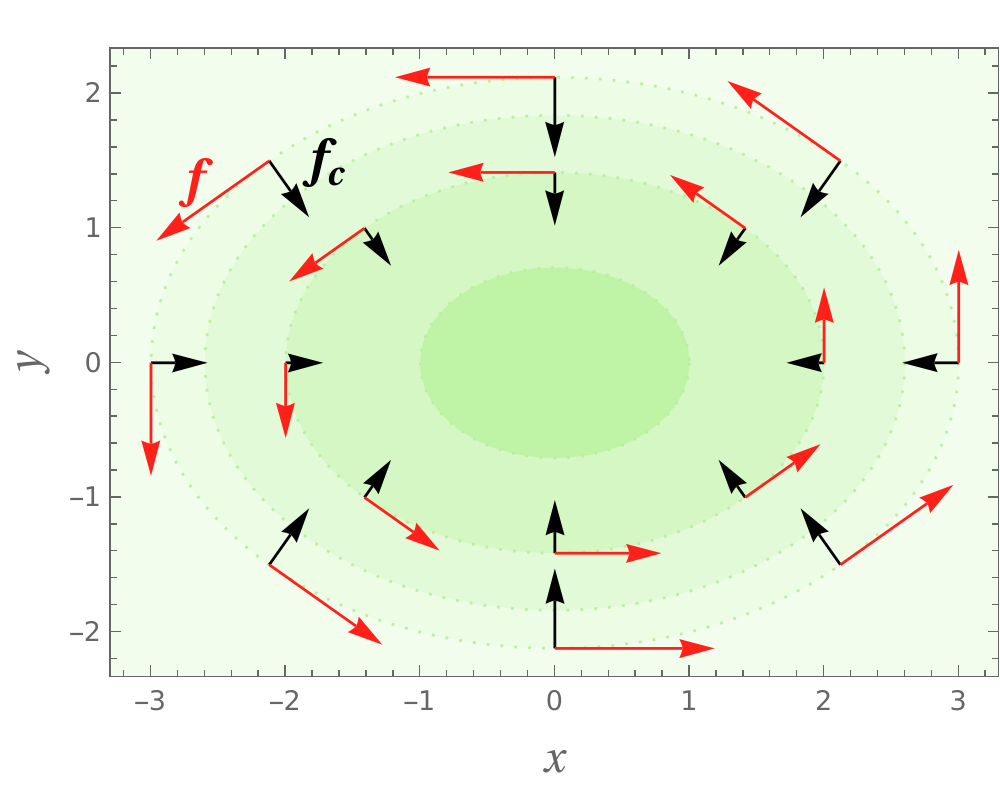}
	\caption{Components of the force field for the Brownian vortex model with $\Phi=2$ and $\alpha=1/2$: black arrows are examples of the conservative component $\pmb{f}_c = -\nabla U$ along equipotential lines, red arrows are the nonequilibrium field $\pmb{f}$ on the same points.}
	\label{fig:BV}
\end{figure}

We consider the Langevin equations \eqref{eq:LE} for a particle with mobility $\mu$ subject to a nonconservative force field in two dimensions. Its position is denoted as $\pmb{x}_{t}=(x_t,y_t)$.
A parabolic potential $U(x,y) = \kappa_{1} x^{2}/2 + \kappa_{2} y^{2}/2 $ contributes with a conservative force $\pmb{f}_e(x_{t},y_{t}) = (-\kappa_{1} x_{t}, -\kappa_{2} y_{t})$ to the dynamics. In addition, in the total deterministic force $\pmb{F}=\pmb{f}_c+\pmb{f}$ there is a non conservative component $\pmb{f}(x,y) = (-\Phi_{1} y, \, \Phi_{2} x)$. The non-conservative and non-reciprocal force $\pmb{f}(x,y)$ is known to drive the system out of equilibrium \cite{loos2020irreversibility,diter24}.  
By setting $\kappa_{1}=\alpha \kappa$, $\kappa_{2}=\kappa$, $\Phi_{1} = \Phi$ and $\Phi_{2} = \alpha \Phi$ here we specialize to a case in which $\pmb{f}_c\perp \pmb{f}$, see the sketch in Figure~\ref{fig:BV}.
Moreover, we take a diagonal diffusion matrix $D_{ij}=k_{B}T \mu \, \delta_{ij}$. The dynamics is thus given by a coupled pair of  linear stochastic differential equations,
\begin{equation}\label{LE_ext_curr}
\begin{split}
\dot x_t & =\mu(-\alpha\kappa\,x_t - \Phi\,y_t)
+ \sqrt{ 2\,k_{B}T\,\mu} ~\xi^{x}_t \\
\dot y_t & = \mu(-\kappa\,y_t + \alpha\Phi \,x_t)
+ \sqrt{ 2\,k_{B}T\,\mu} ~\xi^{y}_t \,,
\end{split}
\end{equation}
with $\langle \xi^{i}_{t'} \rangle = 0$ and $\langle \xi^{i}_{t'} \xi^{j}_{t''} \rangle = \delta_{ij}\, \delta(t'-t'')$.
One can easily verify that the stationary solution of the FP equation has the same Gaussian shape as it would have in equilibrium (i.e~for $\Phi=0$),
\begin{equation}\label{p_st_extcurr}
    p^{\text{st}}_t \sim \exp \left( -\frac{\alpha \,\kappa\,x_{t}^{2}+\kappa\,y_{t}^{2}}{2\,k_{B}T} \right)\, ,
\end{equation}
which implies that $\left< \pmb{x}_{t} \right> =(0,0)$ and
\begin{equation}\label{Cov_ext_curr}
{\cal V}^{ij}_{x} = \frac{k_{B}T}{\alpha \, \kappa}\begin{pmatrix}
1 \hspace{0.5cm}& 0\\[5pt]
0 \hspace{0.5cm}& \alpha
\end{pmatrix}
\end{equation}
where ${\cal V}^{ij}_{x}= \left< x^{i}_{t} \, x^{j}_{t} \right> - \left< x^{i}_{t} \right> \left< x^{j}_{t} \right>=  \left< x^{i}_{t} \, x^{j}_{t} \right>$ is the covariance matrix. Moreover, the mean local velocity defined in \eqref{eq:nu} becomes
\begin{equation}\label{mean_loc_vel_ext_curr}
    \pmb{\nu}_t = \mu\,\left(- \nabla U_t + \pmb{f}_t \right)- k_{B}T \, \mu \,\nabla \ln p^{\text{st}}_t= \mu \, \pmb{f}_t \, ,
\end{equation}
showing that the non-equilibrium state is due to the nonconservative force $\pmb{f}$.

In the following we find analytically the terms of the VSR. 
To calculate relevant quantities associated to the Brownian vortex dynamics, we apply the Laplace transform $\mathcal{L}(g_t) = \hat g_s = \int_0^\infty\mathrm{d}t\, e^{-s t} g_t $ to \eqref{LE_ext_curr},
\begin{equation}
    \begin{cases}
    s \,\hat{x}_s -x_{0}= -\alpha\, \mu\, \kappa\, \hat{x}_s - \mu \,\Phi\, \hat{y}_s + \sqrt{ 2\,k_{B}T\,\mu}~\hat{\xi}^{x}_s\\
    s \,\hat{y}_s -y_{0}= -\mu\, \kappa\, \hat{y}_s + \alpha\, \mu \, \Phi\, \hat{x}_s + \sqrt{ 2\,k_{B}T\,\mu}~\hat{\xi}^{y}_s\\
    \end{cases}
\end{equation}
(where $\mathcal{L}(\dot g_t) = s \hat g_s - g_0$),
which can be rearranged to get 
\begin{equation}\label{ext_curr_sol_step1}
\begin{pmatrix}
\hat{x}_s\\[5pt]
\hat{y}_s
\end{pmatrix}= \hat{\pmb{\chi}}_s \cdot
\begin{pmatrix}
x_{0}+\sqrt{ 2\,k_{B}T\,\mu}~\hat{\xi}^{x}_s\\[5pt]
y_{0}+\sqrt{ 2\,k_{B}T\,\mu}~\hat{\xi}^{y}_s
\end{pmatrix}
\end{equation}
In these equations, written in matrix notation,  we introduced the susceptibility matrix, defined via its Laplace transform,
\begin{equation}
   \hat{\pmb{\chi}}_s =  \frac{1}{(s+\alpha\,\mu\,\kappa)(s+\mu\,\kappa)+ \alpha\,\mu^{2}\,\Phi^{2}} 
\begin{pmatrix}
s+\mu\,\kappa & -\mu \, \Phi \\[5pt]
\alpha\, \mu\, \Phi & s + \alpha \, \mu \, \kappa  
\end{pmatrix} \, .
\end{equation}
By further defining the function
\begin{equation}\label{tau_vortex}
\begin{split}
    \mathcal{T}_t \equiv
    & \mathcal{L}^{-1}\!\left[ \frac{1}{(s+\alpha\,\mu\,\kappa)(s+\mu\,\kappa)+ \alpha\,\mu^{2}\,\Phi^{2}} \right] = \frac{\sin \left( t \,\mu \sqrt{\alpha\,\Phi^2-\frac{(1-\alpha)^{2}\kappa^{2}}{4}} \right)}{\mu\sqrt{\alpha\,\Phi^2-\frac{(1-\alpha)^{2}\kappa^{2}}{4}}}\mathrm{e}^{-\kappa\, \mu(1+\alpha)t/2} \, ,
\end{split}
\end{equation}
one can rewrite the susceptibility matrix,
\begin{equation}
    \pmb{\chi}_t = \begin{pmatrix}
\overset{\,\bm.}{\mathcal{T}}\,{\vphantom{\mathcal{T}}}_t + \mu \, \kappa \mathcal{T}_t & -\mu \, \Phi \,\mathcal{T}_t \\[5pt]
\alpha \, \mu \, \Phi\, \mathcal{T}_t & \overset{\,\bm.}{\mathcal{T}}\,{\vphantom{\mathcal{T}}}_t + \alpha \, \mu \, \kappa \, \mathcal{T}_t
\end{pmatrix} \, ,
\end{equation}
where we also used that $\overset{\,\bm.}{\mathcal{T}}\,{\vphantom{\mathcal{T}}}_t = \mathcal{L}^{-1}\left[s\,\hat{\mathcal{T}}_s \right]$ because $\mathcal{T}(0) = 0$.  With this, by performing an inverse Laplace transform of \eqref{ext_curr_sol_step1} to real time, the solution of \eqref{LE_ext_curr} can be expressed as
\begin{equation}\label{sol_LE_ext_curr}
    x^{i}_t = \chi^{ij}_t \, x^{j}_{0} + \sqrt{2 k_B T\mu}\int_{0}^{t}\mathrm{d}t' \, \chi^{ij}_{t-t'} \, \xi^{j}_{t'} \, ,
\end{equation}
where summation over repeated indexes is understood. In a stationary state with PDF given by \eqref{p_st_extcurr} it holds $\left< \pmb{x}_{t} \right> = (0,0)$ for every $t$ and clearly \eqref{sol_LE_ext_curr} is consistent with this. Moreover, by using \eqref{sol_LE_ext_curr} one can easily calculate the steady-state correlation functions  
\begin{equation}\label{corr_pos_ext_curr}
\begin{split}
    \left< \pmb{x}_{t} \, \pmb{x}^{\text{T}}_{0}\right>  = & \pmb{\chi}_t \left<  \pmb{x}_{0} \,  \pmb{x}^{\text{T}}_{0}\right> = \pmb{\chi}_t {\cal V}_{\pmb{x}} \\[7pt]
    =&\frac{k_{B}T}{\alpha \, \kappa} \begin{pmatrix}
\overset{\,\bm.}{\mathcal{T}}\,{\vphantom{\mathcal{T}}}_t + \mu \, \kappa \mathcal{T}_t & -\alpha\,\mu \, \Phi \,\mathcal{T}_t \\[5pt]
\alpha \, \mu \, \Phi\, \mathcal{T}_t & \alpha\,\overset{\,\bm.}{\mathcal{T}}\,{\vphantom{\mathcal{T}}}_t + \alpha^{2} \, \mu \, \kappa \, \mathcal{T}_t
\end{pmatrix} \, ,
\end{split}
\end{equation}
where we used equation \eqref{Cov_ext_curr} and that, in the Ito convention, $\left<\pmb{\xi}(t')\,\pmb{x}_{0} \right>=0$ for every $t'\ge0$.

The correlation functions shown in \eqref{corr_pos_ext_curr} are the building blocks for covariances of the VSR. We proceed by computing first the covariance matrix of the relative displacement, 
\begin{equation}\label{cov_pos_ext_curr}
\begin{split}
{\cal V}_{\Delta \pmb{x}} (t) =&
\left<
(\pmb{x}_t-\pmb{x}_0)(\pmb{x}^{\mathrm{T}}_t-\pmb{x}^{\mathrm{T}}_0) \right>
-\left<\pmb{x}_t-\pmb{x}_0\right>
\left< \pmb{x}^{\mathrm{T}}_t-\pmb{x}^{\mathrm{T}}_0\right> 
\\=& 
2 \left< \pmb{x}_{0} \, \pmb{x}^{\text{T}}_{0}\right>-\left< \pmb{x}_{t} \, \pmb{x}^{\text{T}}_{0}\right>-\left< \pmb{x}_{t} \, \pmb{x}^{\text{T}}_{0}\right>^{\mathrm{T}}
\\=&
\frac{2\,k_{B}T}{\alpha \, \kappa} \begin{pmatrix}
1- \overset{\,\bm.}{\mathcal{T}}\,{\vphantom{\mathcal{T}}}_t + \mu \, \kappa \mathcal{T}_t & 0 \\[5pt]
0 & \alpha \left( 1- \overset{\,\bm.}{\mathcal{T}}\,{\vphantom{\mathcal{T}}}_t + \mu \, \kappa \mathcal{T}(t) \right)
\end{pmatrix} \, .
\end{split}
\end{equation}
where we used $\left<\pmb{x}_t-\pmb{x}_0\right> =0$.

The covariance matrix of the integral of the forces can be calculated by noting that
\begin{equation}
    F(\pmb{x}_t) = 
    \begin{pmatrix}
    -\alpha\,\kappa\,x_t- \Phi\,y_t \\[3pt]
    -\kappa\,y_t+ \alpha\,\Phi\,x_t
    \end{pmatrix}\, ,
\end{equation}
meaning that the correlation function $ \left< \pmb{F}_{t'} \, \pmb{F}^{\mathrm{T}}_{t''}\right>$ can be again expressed in terms of the components of \eqref{corr_pos_ext_curr}, and that $\left< \pmb{F}_t \right> =0$, which leads to 
\begin{equation}
\label{cov_sum_forc_ext_curr}
\begin{split}
 {\cal V}_{\Sigma_{\pmb{F}}}(t)=&\int_{0}^{t}\mathrm{d}t'\int_{0}^{t}\mathrm{d}t''   \left< \pmb{F}_{t'} \, \pmb{F}^{\mathrm{T}}_{t''}\right>=\\[9pt]
=&\frac{2\,k_{B}T}{ \kappa} \bigintssss_{0}^{t}\mathrm{d}t'\bigintssss_{0}^{t'}\mathrm{d}t'' \begin{pmatrix}
\begin{matrix}
      (\alpha\,\kappa^{2}+\Phi^2)\overset{\,\bm.}{\mathcal{T}}\,{\vphantom{\mathcal{T}}}(t'') + {} \\[2pt]
      +\alpha\, \mu \, \kappa (\kappa^2+\Phi^2) \mathcal{T}(t'')
    \end{matrix}
 & 2\,\Phi\,\kappa (1-\alpha)\overset{\,\bm.}{\mathcal{T}}\,{\vphantom{\mathcal{T}}}(t'') \\[15pt]
2\,\Phi\,\kappa (1-\alpha)\overset{\,\bm.}{\mathcal{T}}\,{\vphantom{\mathcal{T}}}(t'') & \begin{matrix}
      (\kappa^{2}+\alpha\,\Phi^2)\overset{\,\bm.}{\mathcal{T}}\,{\vphantom{\mathcal{T}}}(t'') + {} \\[2pt]
      +\alpha\, \mu \, \kappa (\kappa^2+\Phi^2) \mathcal{T}(t'')
    \end{matrix}
\end{pmatrix} \, .
\end{split}
\end{equation}
Finally, one can calculate the excess variance $\pmb{\EV}(t)$ by applying \eqref{eq:S} and one gets
\begin{equation}
\pmb{\EV}(t) = \frac{ 4\mu^{2}\,k_{B}T}{\kappa} \bigintssss_{0}^{t}\mathrm{d}t'
\begin{pmatrix}
\Phi^{2}\,\mathcal{T}(t') & \Phi \, \kappa \, (1-\alpha)\mathcal{T}(t')\\[3pt]
\Phi \, \kappa \, (1-\alpha)\mathcal{T}(t') & \Phi^{2}\,\mathcal{T}(t')
\end{pmatrix}\, .
\end{equation}

\begin{figure}[t!] 
 	\centering
	\includegraphics[width=0.98\textwidth]{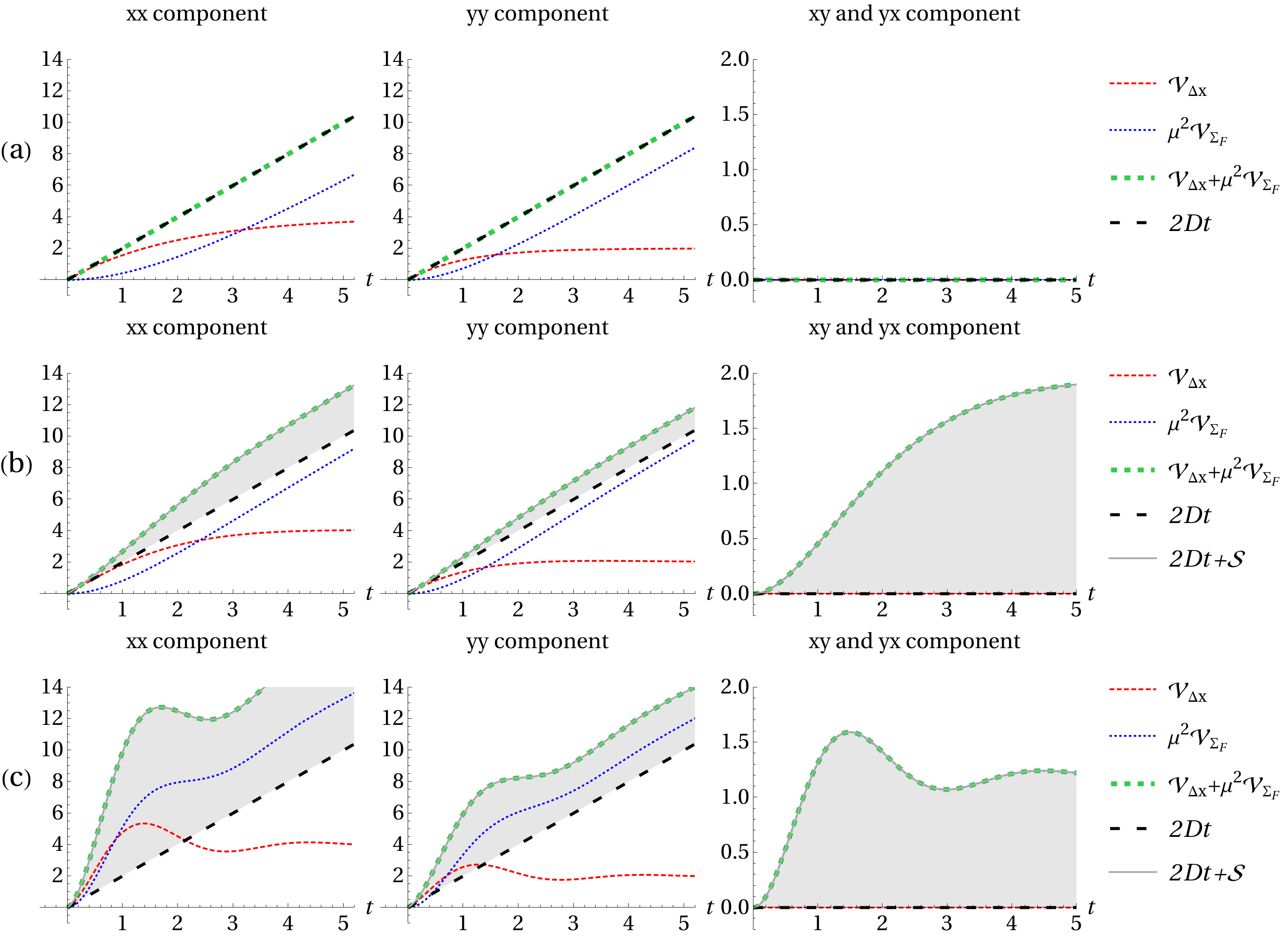}
	\caption{For the Brownian vortex, various quantities in the VSR vs time (see legend, where $\mathcal{V}_{\Sigma_F}$ stands for a component of the covariance of integrated forces in \eqref{cov_sum_forc_ext_curr}); columns of panels are, respectively for $x$-$x$, $y$-$y$, and $x$-$y$ covariances.
	Common parameters are $k_{B}T=\mu=\kappa=1$, $\alpha=1/2$ and different rows are for (a) $\Phi=0$ (equilibrium) (b) $\Phi=3/4$, and (c) $\Phi=3$.
	Each column of panels thus shows the same component of the VSR at different levels of nonequilibrium. Note that $D_{xy}=0$ and $\mathcal{S}_{xy}$ is the only nontrivial term in the $x$-$y$ sector of the VSR.
 The gray areas represent a positive $\mathcal{S}(t)$.}
	\label{fig:VSRvort}
\end{figure}
We have thus obtained all terms in the VSR for the Brownian vortex. As an illustration, in Figure~\ref{fig:VSRvort}, the first row shows examples of the VSR \eqref{eq:VSR} for the Brownian vortex in equilibrium ($\Phi=0$), corresponding to a simple harmonic trap, while the second and third rows show how the excess variance $\mathcal{S}$ (gray area) grows in nonequilibrium conditions and how the sum of position and force variances deviates from $2Dt$ while still fulfilling the VSR.
Since the particle is trapped by the harmonic potential,
 the variance of the displacement converges to a constant at large times (see $xx$ and $yy$ components in the first and second column, respectively).
 Its oscillatory convergence in a strong nonequilibrium regime (Figure~\ref{fig:VSRvort}(c)) is due to the vorticity induced by the nonconservative force $\pmb{f}$.
The asymmetry of the system ($\alpha=1/2$) induces also a non-trivial cross-covariance between the $x$ and the $y$ components (third column of Figure~\ref{fig:VSRvort}).

From the above terms of the VSR we can compute the contributions to the entropy production rate $\sigma$. The diffusion matrix $\pmb{D}$ is diagonal, hence we only need the $x$-$x$ and $y$-$y$ terms:
\begin{subequations}
\begin{align}
\label{sigma_BV_1}
    \frac{1}{4 \mu k_B T} \partial^2_t \mathcal{V}_{\Delta x}|_{t=0} = &
    \frac{\mu  \left(\Phi ^2-\alpha  \kappa ^2\right)}{2 \kappa }
    \\
\label{sigma_BV_2}
    \frac{1}{4\mu k_BT} \partial^2_t \mathcal{V}_{\Delta y}|_{t=0} = &
    \frac{\mu  \left(\alpha  \Phi ^2-\kappa ^2\right)}{2 \kappa }
    \\
\label{sigma_BV_3}
    \frac{\mu}{2 k_BT} \mathcal{V}_{F_x} = &
    \frac{\mu  \left(\Phi ^2 + \alpha  \kappa ^2\right)}{2 \kappa }
    \\
\label{sigma_BV_4}
    \frac{\mu}{2 k_BT} \mathcal{V}_{F_y} = &
    \frac{\mu  \left(\alpha  \Phi ^2+\kappa ^2\right)}{2 \kappa }
    \\
\label{sigma_BV}
    \sigma = & 
    \Phi^{2} \frac{\mu(1+\alpha)}{\kappa} 
\end{align}
\end{subequations}
Here $\sigma$ is computed directly from the second derivative of $\pmb{\EV}(t)$ and, according to our equation \eqref{eq:sigma2}, it is equal to the sum of the terms in \eqref{sigma_BV_1}-\eqref{sigma_BV_4}.
The expression \eqref{sigma_BV} also matches $\sigma$ obtained with Spinney and Ford's formula~\cite{Spinn_ford_entr} using the local velocity $\pmb\nu$ from equation \eqref{mean_loc_vel_ext_curr}, 
\begin{equation}\label{entr_ext_curr}
    \sigma = \big< \pmb\nu_{t} ~ \pmb{D}^{\text{-}1} ~ \! \pmb\nu_{t}  \big> = \frac{\mu \,\Phi^{2}}{k_{B}T\,} \left( \mean{ y_{t}^2}  +\alpha^{2}\mean{ x_{t}^2 } \right) = \Phi^{2} \frac{\mu(1+\alpha)}{\kappa}   \, ,
\end{equation}
where we used $\eqref{corr_pos_ext_curr}$ along with $\left< \pmb{x}_{t}\pmb{x}^{\mathrm{T}}_{t} \right> = \left< \pmb{x}_{0}\pmb{x}^{\mathrm{T}}_{0} \right>$ in this NESS (because $\mean{\pmb{x}_t}=0$), and $\mathcal{T}(0)=0,  \overset{\,\bm.}{\mathcal{T}}\,{\vphantom{\mathcal{T}}}(0)=1$, cf. Eq. \eqref{tau_vortex}. Note that, curiously, $\sigma$ does not depend on the bath temperature in this model.

\begin{figure}[t!] 
 	\centering
	\includegraphics[width=0.85\textwidth]{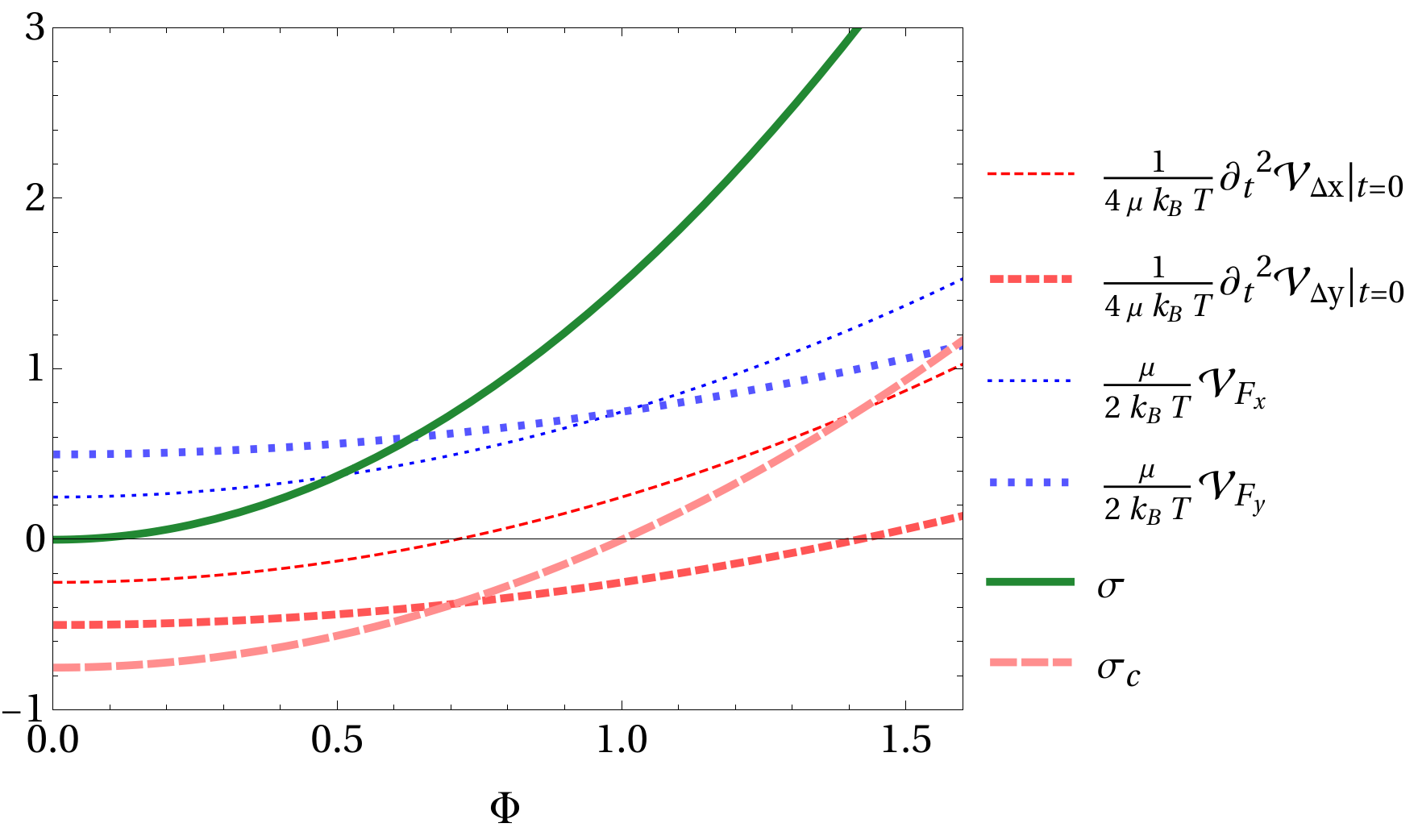}
	\caption{For the Brownian vortex ($T=\mu=\kappa=1$, $\alpha=3$), entropy production rate $\sigma$ vs the strength of the nonequilibrium force $\Phi$, and terms contributing to it.  The sum of the second derivatives of the displacement's covariances, $\sigma_c$, becomes positive at $\Phi=1$. With \eqref{eq:sigma2inf2}, we deem $\Phi>1$ a strong nonequilibrium regime.}
	\label{fig:sigma-vort}
\end{figure}

Figure~\ref{fig:sigma-vort} shows the various terms composing $\sigma$. The illustration shows that pairs of mutually compensating terms \eqref{sigma_BV_1}-\eqref{sigma_BV_3}, and in  \eqref{sigma_BV_2}-\eqref{sigma_BV_4} cancel each other in equilibrium, i.e., at $\Phi=0$.

Empirically, the measurement of the local mean velocity $\pmb\nu$ gives estimates of $\sigma$~\cite{gnesotto2018broken}, even if forces were not measurable.
In addition, one could use \eqref{eq:sigma2inf} and \eqref{eq:sigma2inf2} along with \eqref{sigma_BV_1} and \eqref{sigma_BV_2} to infer $\sigma>0$ and strong nonequilibrium conditions. 
In this model, we know that this occurs for $\Phi>\kappa$, i.e., where the sum of terms with second derivatives turns from negative to positive. It indicates that $\mathcal{V}_{\Delta x}(t)$ and $\mathcal{V}_{\Delta y}(t)$ grow faster than normal diffusion $\sim t$ at short times.

\section{Brownian gyrator}
\label{sec:Br_gy}

The Brownian gyrator is a minimal model for microscopic heat engines operating on the nanoscale \cite{br_gyr_1}. Some of its experimental realisations use circuits with components at different temperatures~\cite{ciliberto2013heat,ciliberto2013statistical,baiesi2016thermal,chiang2017electrical,argun2017experimental}. Following previous theoretical works on this topic \cite{diter24,br_gyr_Villa,br_gyr_2,br_gyr_3,br_gyr_4,Nascimento_2021}, we consider a set of two LEs with a parabolic potential $U(x,y) = \kappa (x^{2}+y^{2})/2+\alpha \,\kappa\,x\,y$, where $-1 < \alpha <1 $ is a factor determining the asymmetry of the potential landscape and with diffusion matrix equal to
\begin{equation}
    \pmb{D} = 
    \begin{pmatrix}
    k_{B}T_{1}\,\mu_{1} & 0\\
    0 & k_{B}T_{1}\,\mu_{2}
    \end{pmatrix} \,
\end{equation}
hence leading to
\begin{equation}\label{LE_brow_gyr}
\begin{cases}
\dot x _t & =  -\mu_1(\kappa\,x_t +\alpha\,\kappa\,y_t) + \sqrt{ 2\,k_{B}T_{1}\,\mu_{1}} ~\xi^{x}_t \\
\dot y _t & = -\mu_2(\kappa\,y_t +\alpha \,\kappa\,x_t) + \sqrt{ 2\,k_{B}T_{2}\,\mu_{2}} ~\xi^{y}_t \, ,
\end{cases}
\end{equation}
with $\langle \xi^{i}_{t'} \rangle = 0$ and $\langle \xi^{i}_{t'} \xi^{j}_{t''} \rangle = \delta_{ij}\, \delta(t'-t'')$. The solution of this model is in~\ref{app:BG}.
In the following, we show examples of its VSR and discuss the terms contributing to entropy production.

\begin{figure}[t!] 
 	\centering
\includegraphics[width=0.98\textwidth]{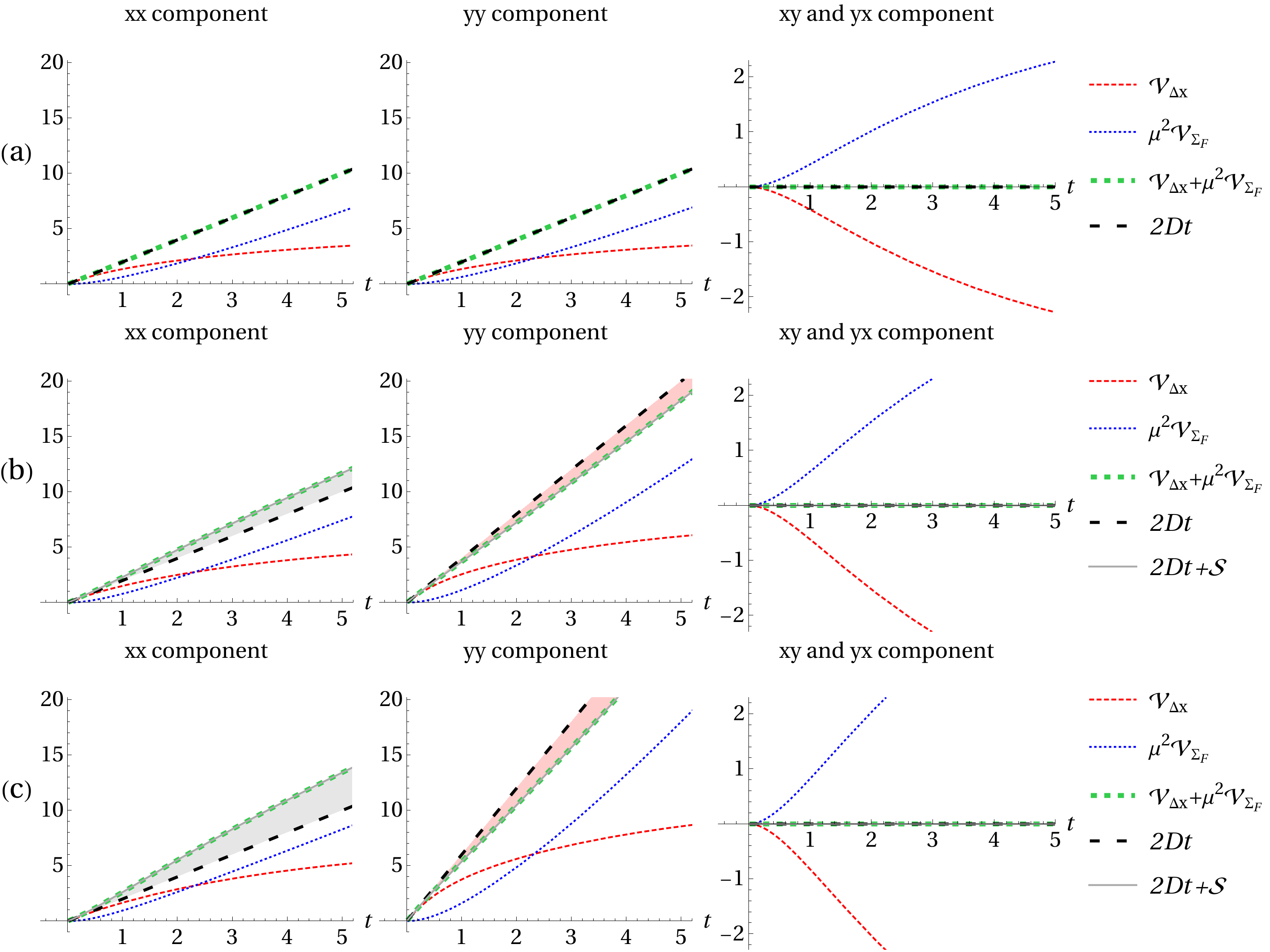}
\caption{For the Brownian gyrator, various quantities in the VSR vs time (see the legend, where $\mathcal{V}_{\Sigma_F}$ stands for a component of the covariance of integrated forces in \eqref{cov_sum_forc_br_gyr}); columns of panels are, respectively for $x$-$x$, $y$-$y$, and $x$-$y$ covariances. Common parameters are $k_{B}T_1=\mu_1=\mu_2=\kappa=1$, $\alpha=3/4$ and different rows are for (a) $T_2=T_1$ (equilibrium), (b) $k_{B}T_2=2$, and (c) $k_{B}T_2=3$.
 For the $x$-$x$ component, the grey area represents a positive $\mathcal{S}(t)$ while, for the $y$-$y$ component, the pink area represents a negative $\mathcal{S}(t)$.}
\label{fig:VSRgyr}
\end{figure}

In Figure~\ref{fig:VSRgyr} we show an example of VSR for all components of the Brownian gyrator. In this case, there emerges a novel behaviour for the excess variance $\mathcal{S}(t)$: it turns out to be negative for the component $y$-$y$, which is the degree of freedom in the reservoir at higher temperature $T_2>T_1$.

From the various (excess) variances found in~\ref{app:BG}, we can compute the contributions to $\sigma$ \eqref{eq:sigma2}. As for the Brownian vortex, since the diffusion matrix $\pmb{D}$ is diagonal, we just need the $x$-$x$ and $y$-$y$ contributions to $\sigma$,
\begin{subequations}
\begin{align}
\label{sigma_BG_1}
    \frac{1}{4\mu_1 k_B T_1} \partial^2_t \mathcal{V}_{\Delta x}|_{t=0} = &
    \frac{\alpha^2\kappa  \mu_1  \mu_2 (T_2-T_1)}{2 (\mu_1+\mu_2) T_1}-\frac{\kappa\mu_1}{2}
    \\
\label{sigma_BG_2}
    \frac{1}{4\mu_2 k_BT_2}  \partial^2_t \mathcal{V}_{\Delta y}|_{t=0} = &
    \frac{\alpha^2\kappa  \mu_1  \mu_2 (T_1-T_2)}{2 (\mu_1+\mu_2) T_2}-\frac{\kappa\mu_2}{2}
    \\
\label{sigma_BG_3}
    \frac{\mu_1}{2 k_BT_1} \mathcal{V}_{F_x} = &
    \frac{\alpha^2\kappa  \mu_1  \mu_2 (T_2-T_1)}{2 (\mu_1+\mu_2) T_1}+\frac{\kappa\mu_1}{2}
    \\
\label{sigma_BG_4}
    \frac{\mu_2}{2k_B T_2} \mathcal{V}_{F_y} = &
    \frac{\alpha^2\kappa  \mu_1  \mu_2 (T_1-T_2) }{2 (\mu_1+\mu_2) T_2}+\frac{\kappa\mu_2}{2}
\end{align}
\begin{align}
\label{sigma_BG_5}
    \sigma_x = & 
    \eqref{sigma_BG_1} + \eqref{sigma_BG_3} 
    =
    \alpha^2 \kappa \frac{  \mu_1 \mu_2 (T_2-T_1)}{(\mu_1+\mu_2)T_1}
    \\
\label{sigma_BG_6}
    \sigma_y = & 
    \eqref{sigma_BG_2} + \eqref{sigma_BG_4} 
    =
    \alpha^2 \kappa \frac{ \mu_1 \mu_2 (T_1-T_2)}{(\mu_1+\mu_2)T_2}
    \\
\label{sigma_BG}
    \sigma = &
    \sigma_x + \sigma_y =
    \alpha^2 \kappa  \frac{\mu_1 \mu_2 }{\mu_1+\mu_2} \frac{(T_1-T_2)^2}{ T_1 T_2} \, .
\end{align}
\end{subequations}
From this expression, one can note that $\sigma$ becomes equal to zero if global equilibrium with $T_{1}=T_{2}$ is established or if the potential energy has no cross term ($\alpha=0$), preventing energy transfer between $x$ and $y$ components, which thermalise independently.

\begin{figure}[t!] 
 	\centering
\includegraphics[width=0.85\textwidth]{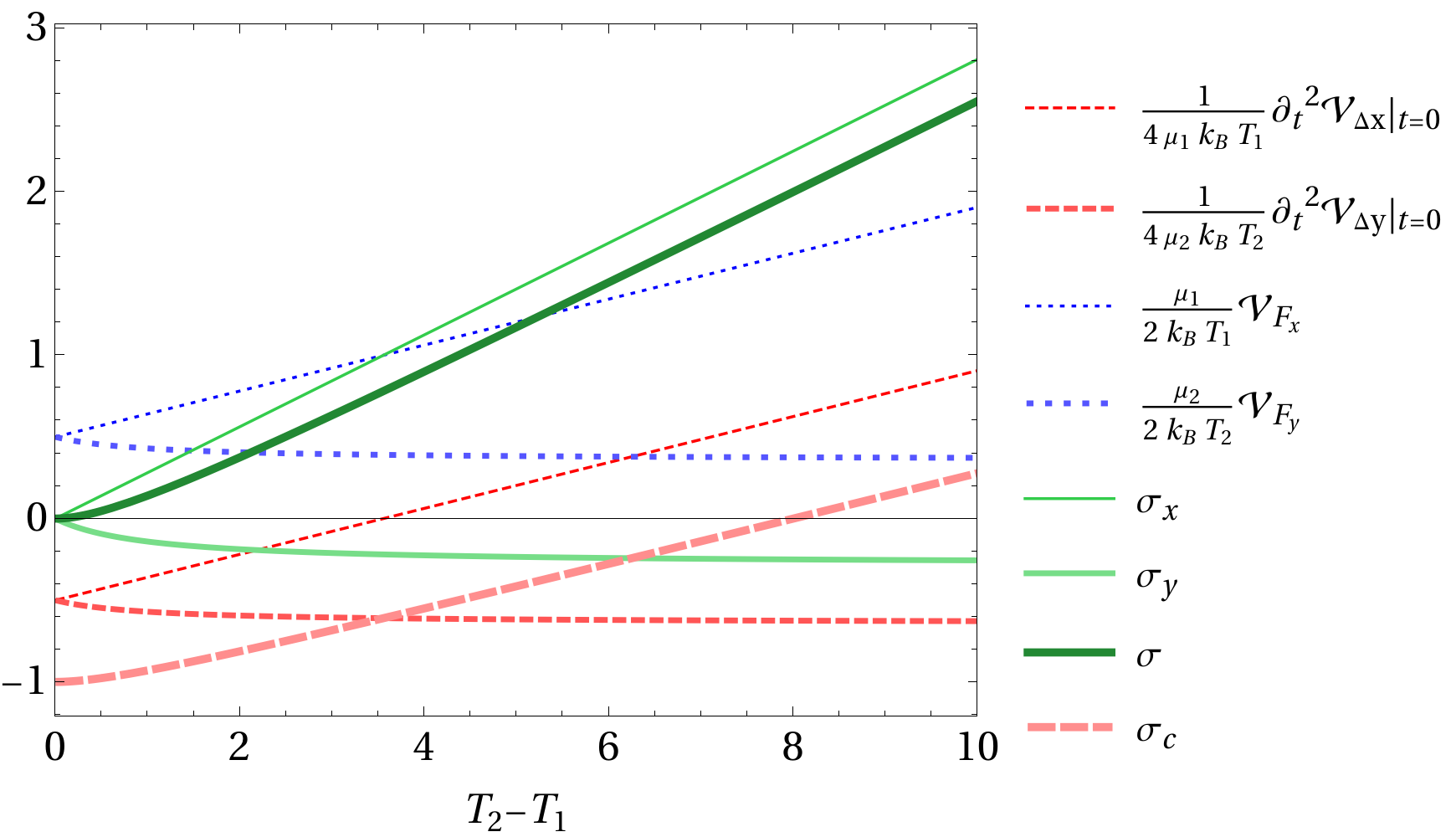}
	\caption{For the Brownian gyrator, we plot various terms contributing to $\sigma$ and its total value as a function of the temperature imbalance $T_2-T_1$. According to \eqref{eq:sigma2inf2}, strong nonequilibrium conditions emerge where the sum of the second derivatives of the displacement's covariances, $\sigma_c$, becomes positive.}
	\label{fig:sig-gyr}
\end{figure}

In Figure~\ref{fig:sig-gyr} we plot all terms above as a function of the temperature difference $T_2-T_1$. Due to energy equipartition, \eqref{sigma_BG_2} is equal to \eqref{sigma_BG_4} in equilibrium, where also \eqref{sigma_BG_1}=\eqref{sigma_BG_3} for symmetry.
By increasing the difference $T_2-T_1$, we recover that the degree of freedom at the higher temperature ($y$) extracts on average heat from its reservoir ($T_2 \sigma_y<0$). The system delivers it to the colder reservoir ($T_1 \sigma_x>0$), with total entropy production rate $\sigma>0$. Again, we note a point $T_2-T_1\simeq 8$ where the sum of second time-derivatives \eqref{sigma_BG_1} and \eqref{sigma_BG_2} of the displacement variances (${\cal V}_{\Delta x}(t)$ and ${\cal V}_{\Delta y}(t)$) become positive, marking the onset of the strong non-equilibrium regime where $\sigma_c \ge 0$ (cf. Eq. \eqref{eq:sigma2inf}).

\section{Discussion}
\label{sec:disc}

We have studied the recently introduced VSR \cite{VSR}, which helps visualise the degree of nonequilibrium in overdamped diffusive systems and leads to novel formulas for the entropy production rate $\sigma$. In dimensionless units, our main formula \eqref{eq:sigmaS} for $\sigma$ reduces, for diagonal diffusion matrices, to \eqref{eq:sigma2}, i.e.,
\begin{equation}
\tag{\ref{eq:sigma2}}
    \sigma 
    ~= 
    \sum_{i=1}^d
    \frac 1 {k_B T_i}
    \left[
    \frac{1}{\mu_i} ({v}^i)^2 
    +
    \left. 
    \frac{1}{4 \mu_i}  \partial_{t}^{2}
    \mathcal{V}^{i}_{\Delta x}(t)
    \right|_{t=0}
    + 
    \frac{\mu_i}{2}
    \mathcal{V}^{i}_{F} 
    \right]
    \,.
\end{equation}
For the first time, \eqref{eq:sigma2} shows that the second-order time derivative of the displacement's variance contains information on the dissipative processes underlying the particles' dynamics. The sum of these derivatives is negative in equilibrium to compensate for the positive sum of force variances. This leads us to introduce a criterion for discriminating strong nonequilibrium conditions, marked by the point where the sum of second-order time derivatives of displacement's variances, $\sigma_c$, becomes positive, see~\eqref{eq:sigma2inf} and~\eqref{eq:sigma2inf2}. Equation \eqref{eq:sigma2} and the more general form \eqref{eq:sigma_VSR} with their second-order time derivative term at time zero bear resemblance with the short-time thermodynamic uncertainty relation derived by Manikandan and collaborators \cite{manikandan2018exact,manikandan2020inferring,manikandan2021quantitative}. In this case, currents must be optimized to find good estimates; however, in most cases, $\sigma$ values are too small \cite{manikandan2022estimate,lynn2021broken}. 

An intermediate step in our derivation is
\eqref{eq:EP_sekimoto_final}, which here is simplified for diagonal diffusion matrices to
\begin{equation}
\sigma = 
    \sum_{i}\frac{1}{k_B T_i}\left[
    v^{i}\mean{F^i} +
    \frac{\overset{\,\bm .}{C}\,{\vphantom{C}}_{\!xF}^{i}(0^+)-\overset{\,\bm .}{C}\,{\vphantom{C}}_{\!Fx}^{i}(0^+)}{2}\right] \,
\label{eq:new-int}
\end{equation}
This formula stems from the standard one for entropy production in a NESS, derived in stochastic energetics~\cite{sek10},
\begin{equation}
\sigma= \sum_{i}\frac{1}{k_B T_i}\bigg\langle F^i \circ \frac{{dx}^{i}}{dt}\bigg\rangle
\tag{\ref{eq:EP_sekimoto}}
\end{equation}
for which Sekimoto~\cite{sek10} also provided an alternate version,
\begin{equation}
\sigma= \sum_{i} \frac{\mu_i}{k_B T_i}
\left[ \Mean{(F^i)^2} + k_B T_i
\Mean{\frac{\partial F^i}{\partial x_i}}
\right]
\label{eq:EP_sekimoto_2}
\end{equation}
We may rewrite \eqref{eq:sigma2} to get closer to \eqref{eq:EP_sekimoto_2}.
Since the the average velocity $v^i = \mean{x^i_t-x^i_0} / t$ is related to the average force by $v^i = \mu_i \mean{F^i}$, one can reshape the variance of forces $\mathcal{V}^{i}_{F}  = \mean{(F^i)^2} - \mean{F^i}^2 = \mean{(F^i)^2} - (v^i/\mu_i)^2$ in \eqref{eq:sigma2}. Furthermore, since the variance of the displacement is related to the autocorrelation function ($\mathcal{V}^{i}_{\Delta x}(t) = \mean{(x^i_t)^2}+\mean{(x^i_0)^2}- 2 \mean{x^i_tx^i_0}-\mean{x^i_t-x^i_0}^2$), we can rewrite
\eqref{eq:sigma2} as
\begin{equation}
\label{eq:sigma3}
    \sigma 
    ~= 
    \sum_{i=1}^d
    \frac {\mu_i} {2 k_B T_i}
    \left[
    \left. \left(\frac{{v}^i}{\mu_i}\right)^2+
    \Mean{(F^i)^2}
    -\frac{1}{\mu_i^2}  \partial_{t}^{2}
    C_{xx}^{i}(t)
    \right|_{t=0}
    \right]
    \,.
\end{equation}
This formula is our last new result. Note the presence of a $1/2$ prefactor in~\eqref{eq:sigma3}, which is not present in~\eqref{eq:EP_sekimoto_2}.

All formulas above rely on measurements of forces, and \eqref{eq:EP_sekimoto} is the simplest among them and does not need mobility values. The formula \eqref{eq:EP_sekimoto_2} is purely based on instantaneous averages but requires the knowledge of the forces' gradients, which might be nontrivial to obtain. Knowledge of forces can lead to good estimates of $\sigma$ \cite{li2019quantifying}, but an experimental challenge is sampling stochastic trajectories with a fast enough rate of $1/dt$. Indeed, a slow sampling rate could bias the estimate of the Stratonovich integration in \eqref{eq:EP_sekimoto}, making it less appealing than the other formulas involving instantaneous averages, which do not depend on the sampling rate. For example, a numerical simulation of the Brownian vortex model of Sec.\ref{sec:BV} shows that the precision of the VSR in predicting $\sigma$ is higher than using \eqref{eq:EP_sekimoto}. This is apparent upon decreasing the sampling rate, which is equal to the inverse of the simulation time-step \cite{VSR_computations_Inprep}. In particular, the precision of the estimate using the Stratonovich product \eqref{eq:EP_sekimoto} underestimates the true value by 20$\%$ when the sampling rate decreases by one order of magnitude beyond the system's relaxation time. In contrast, the value of $\sigma$ obtained with the VSR remains insensitive to the sampling rate.  Consequently, the VSR might be useful in numerical simulations with limited computational resources, where a low sampling rate is needed to simulate traces over sufficiently long times. Similar considerations apply to the analysis of experimental data recorded at low sampling rates.

Our novel \eqref{eq:sigma2} and \eqref{eq:sigma3} replace estimates of forces' gradients with computations of second-order time derivatives. Moreover, since $\mu_i$ enters both their numerators and denominators, they are sensible to the values of mobilities. They can thus be used in conjunction with the other formulas for $\sigma$ to perform solid parameter estimations.

If forces cannot be measured, one could estimate local mean velocities $\pmb{\nu}(\pmb x)$ in order to use 
\begin{equation}
    \sigma = \big< \pmb\nu ~ \pmb{D}^{\text{-}1} ~ \! \pmb\nu  \big>
\end{equation}
for measuring the entropy production rate~\cite{Spinn_ford_entr,gnesotto2018broken}. Such an approach does not work for confined one-dimensional signals, such as optical tweezers measurements of flickering red blood cells~\cite{turlier2016equilibrium,VSR}.
In this case, one can use the Harada-Sasa formula to compute $\sigma$ without measuring forces. However, it needs the computation of the response $R_{xx}^i$ to linear forces, eventually performed with active measurements by space-independent oscillatory forcing in Fourier space at all frequencies $\omega/2\pi$,
\begin{equation}
\label{eq:HS}
    \sigma 
    ~= 
    \sum_{i=1}^d
    \frac {1} {k_B T_i \mu_i}
    \left\{
    ({v}^i)^2 + 
    \int_{-\infty}^{\infty}
    \frac{d\omega}{2\pi}
    \left[
    \tilde C_{xx}^{i}(\omega)
    - \mathrm{Re}\left(
    \tilde R_{xx}^{i}(\omega)\right)
    \right]
    \right\}
    \,.
\end{equation}
If the response of the system cannot be measured, the last solution is replacing it by some modeling. Indeed, the reduced version of our VSR~\cite{VSR} can detect and quantify entropy production from one-dimensional signals.

The VSR sets a resource for deriving entropy production rates in non-equilibrium systems. It relies on an equality between the sum of the variances of displacement and impulses. Whenever positions and forces are measurable, one can extract the excess variance to derive $\sigma$ using~\eqref{eq:sigmaS}. The result should not be different from what can be obtained by directly measuring $\sigma$ using the standard formula from stochastic energetics involving the time-average of a Stratonovich product between forces and velocities, \eqref{eq:EP_sekimoto}.

If forces are only partially measurable, then the VSR cannot be directly applied. To overcome this problem, we have introduced in Ref.\cite{VSR} a reduced-VSR that combines the measured forces with those that remain inaccessible, using a theoretical model for the latter. The reduced VSR yields a new equality similar to \eqref{eq:VSR} relying only on the impulses of the measurable forces on the left-hand side, and a new model-dependent excess variance $\widetilde{S}$ in the right-hand side. For practical applications where the knowledge of forces is limited, a model prediction is fitted to the reduced VSR.  The tight constraint imposed by the reduced VSR over the experimental timescales, typically spanning several decades, gives robust predictions of $\sigma$, linking modeling to energetics. Therefore, exactly solvable models are key to deriving $\sigma$ from experiments. Whenever models are not solvable, numerical approaches can be useful. Examples are methods based on machine-learning \cite{otsubo2020estimating,kim2020learning}, simulation-based inference (SBI) \cite{SBI_PNAS,Tucci22}, and reconstruction of force-fields \cite{perez2018high,frishman2020learning,Argun20forcefields}. Exactly solvable models combined with the VSR find applications in cases where the accessible experimental variables have Gaussian statistics hiding nonequilibrium behavior and $\sigma$ \cite{br_gyr_Villa,doerries2021correlation,lucente2022inference,netz2023multi}. The vast amount of existing work on stochastic thermodynamics models and the large amount of available experimental data in systems at the nanoscale foresee new applications of the VSR.

\paragraph{Acknowledgements} MB thanks the MPI PKS in Dresden for the hospitality. IDT thanks the INFN Section in Padova for the hospitality. FR thanks Icrea Academia Prizes 2018 and 2023, and is supported by the Spanish Research Council (grant PID2019-111148GB-100 and PID2022-139913NB-100).

\appendix

\section*{Appendices}

The first three sections of the appendix show how to express the excess variance as a function of the mean local velocity. First, we derive two results necessary for that proof: we express a response function in terms of the mean local velocity (\ref{app:R}), and we prove that there is no correlation between the position measured in the frame moving with the mean local velocity, and the mean local velocity itself (\ref{app:y}). Finally,  \ref{app:Deriv_vel} derives the excess variance.

In \ref{app:BG}, we report details on the analytical solution of the Brownian gyrator model.

\section{Response function}
\label{app:R}

For dynamics following the overdamped Langevin equation
\begin{equation}
\tag{\ref{eq:LE}}
\begin{split}
     \dot{\pmb{x}}_t &
     =  \pmb{\mu} \, \pmb{F}_{t} + \sqrt{2\,\pmb{D}} \cdot \pmb{\xi}_t
\end{split}
\end{equation}
we aim to compute the response function
\begin{equation}\label{resp_func}
    \mathcal{R}^{ij}(t,t') = \bigg \langle \frac{\delta ~\! x^{i}_{t} }{\delta\xi^{j}_{t'}} \bigg \rangle \,.
\end{equation}
and prove
\begin{align}\label{Rik}
\sqrt{2 ~\!D_{jk}} ~\!\mathcal{R}^{ik}(t,t') 
= 2 ~\!\bigg( \frac{\mathrm{d}~\!}{\mathrm{d}~\! t'} \langle x^{i}_{t}~ x^{j}_{t'}\rangle -  \langle x^{i}_{t} ~ \nu^{j}_{t'} \rangle \bigg)
\end{align}
with response theory. Eq.~\eqref{Rik} reproduces linear response relations based on the mean velocity $\pmb{\nu}$~\cite{spe06,che08} and will be useful in~\ref{app:Deriv_vel} for expressing the excess variance as a function of $\pmb{\nu}$.

If a perturbing potential $-h V(x) $ is added to the system so that the total force becomes
\begin{equation}\label{H_pert}
    ({F^{tot}_t})^j = F^j_t + f^j_t = F^j_t + h_t ~ \partial_{x_j} V_t \, ,
\end{equation}
a formula for the response in the \textit{steady state} of an observable $\mathcal{O}(t)$ to an instantaneous variation of $h_{t'}$ is~\cite{bai09,bai09b,bai11}
\begin{equation} \label{FDT_gen}
\begin{split}
     \bigg \langle \frac{\delta \mathcal{O} _{t} }{\delta h^{j}_{t'}} \bigg \rangle =&~ (k_{B}\pmb{T})_{jl}^{\textrm{-}1}\bigg(\frac{\mathrm{d}~}{\mathrm{d} t'} \langle \mathcal{O}_{t}V^l_{t'} \rangle - \frac{1}{2} \langle \mathcal{O} _{t} ~\! (L_{t'}-L^{*}_{t'}) V^l_{t'} \rangle-\frac{1}{2}\langle \mathcal{O} _{t} ~ \partial_{t'}~\! V^l_{t'} \rangle \bigg) \,,
\end{split}
\end{equation}
where $T_{jl}$ is the diagonal temperature matrix and $L_{t'}$ and $L^{*}_{t'}$ are, respectively, the generator of the dynamics and its conjugate. For the systems we are considering, they are equal to
\begin{equation}
    L_{t} = \mu_{ik} ~\! F^{i}_{t} ~\! \partial_{x^{k}_{t}} +  D_{ik} ~\! \partial_{x^{i}_{t}} \partial_{x^{k}_{t}} 
\end{equation}
\begin{equation}
    L^{*}_{t} = -\mu_{ik} ~\! F^{i}_{t}  \partial_{x^{k}_{t}} -  D_{ik} ~\! \partial_{x^{i}_{t}}  \partial_{x^{k}_{t}} + 2~\! D_{ik} ~\! \big( \partial_{x_{t}^{i}} \log(p^{\text{st}}(\pmb{x}_t)\big)   \partial_{x^{k}_{t}}
\end{equation}
where Einstein's notation is understood and  $p^{\text{st}}(\pmb{x}_t,t)$ is the stationary distribution obtained from the FP equation associated to \eqref{eq:LE}.
The same formal reasoning that leads to \eqref{FDT_gen} can be applied to the thermal noise, which near equilibrium leads to the famous Onsager regression hypothesis. Indeed, we see that the Langevin equation can be rewritten as
\begin{equation}
    \dot x^{i}(t) = \mu_{ij} ~\! F^{j}_{t} + \sqrt{2~\!D_{ij}}~\xi^{j}_t = \mu_{ij} ~ \! \big( F^{j}_{t} + \mu^{\text{-}1}_{jl}~\! \sqrt{2~\!D_{lr}} ~ \xi^{r}_t \big) \, ,
\end{equation}
meaning that the thermal noise can be seen as a small perturbation to the deterministic dynamics driven by $\pmb{F}_{t}$ and arising from the perturbing force
\begin{equation}
    f^j_t =  \mu^{\text{-}1}_{jl}~\! \sqrt{2~\!D_{lr}} ~ \xi^{r}_t  \, .
\end{equation}
Moreover, comparing this last equation with \eqref{H_pert}, we identify
\begin{align}\label{pert_pot}
    V_t &=  x^{r}_t~\! \mu^{\text{-}1}_{rl}~\! \sqrt{2~\!D_{lj}}\\
    h_t &= \xi^{j}_t
\end{align}
so that \eqref{FDT_gen} specialises to 
\begin{equation} 
\label{resp_O_xi}
     \bigg \langle \frac{\delta \mathcal{O} _{t} }{\delta \xi^{j}_{t'}} \bigg \rangle =(k_{B}\pmb{T})_{js}^{\textrm{-}1}\sqrt{2~\!D_{sl}}~ \mu^{\text{-}1}_{lr} \left( \frac{\mathrm{d}~\!}{\mathrm{d}~\! t'} \langle \mathcal{O}_{t}\,x^{r}_{t'}\rangle -  \langle \mathcal{O} _{t} ~\! (L_{t'}-L^{*}_{t'})~  x^{r}_{t'} \rangle/2\right)
\end{equation}
as a consequence of $\mu_{ij}$ and $D_{ij}$ being symmetric and $\partial_{t} V_t  =0 $, because $\partial_{t}$ is a partial derivative and $V_t$ in \eqref{pert_pot} does not depend explicitly on time. 
The response function \eqref{resp_func}, for an observable $\mathcal{O}=x^i$, with \eqref{resp_O_xi} becomes
\begin{equation}\label{resp_2}
    \mathcal{R}^{ij}(t,t') 
    =  (k_{B}\pmb{T})_{js}^{\textrm{-}1}\sqrt{2~\!D_{sl}}~ \mu^{\text{-}1}_{lr}~\! \left( \frac{\mathrm{d}~\!}{\mathrm{d}~\! t'} \langle x^{i}_{t}~ x^{r}_{t'}~ \! \rangle -  \langle x^{i}_{t} ~\! (L_{t'}-L^{*}_{t'})~\!  x^{r}_{t'} ~\! \rangle/2 \right)\, .
\end{equation}
Let us calculate the term involving the generators of the dynamics, i.e.
\begin{equation}\label{operators}
\begin{split}
    (L_{t'}-L^{*}_{t'})  x^{r}_{t'}  =& 2~\! \Big( \mu_{kl} ~\! F^{k}_{t'}\,  \partial_{x^{l}_{t'}} +  D_{kl} ~\! \partial_{x^{k}_{t'}} \partial_{x^{l}_{t'}} - D_{kl} ~\! \big( \partial_{x^{k}_{t'}} \log(p^{\text{st}}_{t'}\big)  \partial_{x^{l}_{t'}} \Big) ~\!  x^{r}_{t'}\\
    =& 2 ~\! \Big( \mu_{kr} ~\! F^{k}_{t'}  - D_{kr} ~\! \partial_{x^{k}_{t'}} \log (p^{\text{st}}_{t'}) \Big) = 2~\! \nu^{r}_{t'}  \, , 
\end{split}
\end{equation}
where $\nu^{r}_{t}$ is a component of the steady state mean local velocity defined in \eqref{eq:nu}. By plugging equation \eqref{operators} into \eqref{resp_2} and relabeling indexes, one obtains that 
\begin{equation}\label{resp_3}
    \mathcal{R}^{ij}(t,t') =  (k_{B}\pmb{T})_{js}^{\textrm{-}1} \sqrt{2~\!D_{sl}}~ \mu^{\text{-}1}_{lr}~\! \bigg( \frac{\mathrm{d}~\!}{\mathrm{d}~\! t'} \langle x^{i}_{t}~ x^{r}_{t'}\rangle -  \langle x^{i}_{t} ~ \nu^{r}_{t'} \rangle \bigg) \, . 
\end{equation}
Hence, one can also note that 
\begin{align}\label{inter1_ch5}
\sqrt{2 ~\!D_{jk}} ~\!\mathcal{R}^{ik}(t,t') =& \sqrt{2 ~\!D_{jk}}~\!(k_{B}\pmb{T})_{ks}^{\textrm{-}1} \sqrt{2~\!D_{sl}}~ \mu^{\text{-}1}_{lr}~\! \bigg( \frac{\mathrm{d}~\!}{\mathrm{d}~\! t'} \langle x^{i}_{t}~ x^{r}_{t'}\rangle -  \langle x^{i}_{t} ~ \nu^{r}_{t'} \rangle \bigg)\nonumber\\
    =& \sqrt{2 ~\!D_{jk}} \sqrt{2~\!D_{ks}}~\!(k_{B}\pmb{T})_{sl}^{\textrm{-}1} ~ \mu^{\text{-}1}_{lr}~\! \bigg( \frac{\mathrm{d}~\!}{\mathrm{d}~\! t'} \langle x^{i}_{t}~ x^{r}_{t'}\rangle -  \langle x^{i}_{t} ~ \nu^{r}_{t'} \rangle \bigg)\nonumber\\
    =& 2 ~\!D_{js}~\! D^{\text{-}1}_{sr}~\! \bigg( \frac{\mathrm{d}~\!}{\mathrm{d}~\! t'} \langle x^{i}_{t}~ x^{r}_{t'}\rangle -  \langle x^{i}_{t} ~ \nu^{r}_{t'} \rangle \bigg)\nonumber\\
    =& 2 ~\!\bigg( \frac{\mathrm{d}~\!}{\mathrm{d}~\! t'} \langle x^{i}_{t}~ x^{j}_{t'}\rangle -  \langle x^{i}_{t} ~ \nu^{j}_{t'} \rangle \bigg) \, ,
\end{align}
where between the first and the second line, we used that, if $\pmb{T}$ and its inverse are diagonal (but not proportional to the identity), also $\pmb{D}$ and $\pmb{\mu}$ are diagonal and hence all matrices involved in the previous formula commute. Instead, if $\pmb{T}$ is proportional to the identity, it trivially commutes with all matrices. For the same reason, we have that $\pmb{D}^{\mathrm{-}1}=(k_{B}\pmb{T}~\!\pmb{\mu})^{\mathrm{-}1}=k_{B}^{\mathrm{-}1}\pmb{T}^{\mathrm{-}1}~\!\pmb{\mu}^{\mathrm{-}1}$, motivating the step between the second and the third line of \eqref{inter1_ch5}. 

\section{Null covariance between position and mean local velocity}
\label{app:y}

In a steady state with nonzero mean drift velocity $\pmb{v} = \langle \pmb{\nu} \rangle$ and constant diffusion matrix, we may write the position
$\pmb{x}_t = \pmb{v}t + \pmb{y}_t$ as a sum of a deterministic component $\pmb{v}t$ and a random component $\pmb{y}_t$, whose average $\mean{\pmb{y}_t} = \mean{\pmb{y}_0} = \mean{\pmb{y}}$ is constant so that $\langle \pmb{x}_{t} \rangle = \pmb{v}t+ \langle \pmb{y} \rangle$. This is indeed a consequence of Galilean invariance applied to the LE \cite{Cairoli_Galileian_invariance}. Therefore, the steady-state averages involving $\pmb{x}$, and including terms scaling linearly with time, are grounded on those as a function of $\pmb{y}$. The same is true for two-time correlations. Moreover, taking averages at time zero is convenient because $\pmb{x}_0=\pmb{y}_0$.

We show that there is zero covariance between position and mean local velocity:
\begin{align}
\label{zero_cov}
\left[\mathrm{Cov}(y^i_t \,,\nu^j_t)\right]\sym = 
\left[\mathrm{Cov}(y^i_0 \,,\nu^j_0)\right]\sym = 
\left[\mathrm{Cov}(x^i_0 \,,\nu^j_0)\right]\sym = 0
\end{align}
where, e.g.,
\begin{align}
\left[\mathrm{Cov}(x^i_0 \,,\nu^j_0)\right]\sym
&=
    \left[C^{ij}_{ x \nu }(0) \right]\sym = \frac{C^{ij}_{x \nu }(0) + C^{ji}_{x \nu }(0)}{2}
    =
    \nonumber\\
    &=
    \frac 1 2
    \left(
    \mean{ x_0^i \nu_0^j } 
    -\mean{ x_0^i}\mean{\nu_0^j }
    + \mean{ x_0^j \nu_0^i } 
    -\mean{ x_0^j}\mean{\nu_0^i }
    \right)
\end{align}
This result will be used in the last steps of the derivation in~\ref{app:Deriv_vel}.

First, using the definition \eqref{eq:nu} of the local mean velocity, we note that
\begin{align}\label{proof_B1}
\left[\mean{x^{i}_{0} ~ \nu^{j}_{0}}\right]\sym
=& \left[\mu_{jk}\langle x^{i}_{0} ~ F^{k}_{0} \rangle\right]\sym 
- \! \int \!\mathrm{d}\pmb{x}_{0}~\!
\left[ D_{jk}~\!\partial_{x^{k}_{0}} p^{\text{st}}(\pmb{x}_{0})x^{i}_{0} \right]\sym
\nonumber\\
=& \left[\mu_{jk}\langle x^{i}_{0} ~ F^{k}_{0} \rangle\right]\sym 
+ \! \int \!\mathrm{d}\pmb{x}_{0}~\!
\left[D_{ji}~\! p^{\text{st}}(\pmb{x}_{0}) \right]\sym
\nonumber\\
=&\left[\mu_{jk}\langle x^{i}_{0} ~ F^{k}_{0} \rangle\right]\sym +\!D_{ij}\, .
\end{align}
The next step involves equating
\begin{align}\label{proof_B2a}
\langle x^{i}_{dt}  x^{j}_{dt} \rangle 
&= \langle x^{i}_{0}  x^{j}_{0} \rangle + 2 \left[\mu_{ik} \mean{x_0^i F_0^k}\right]\sym dt + 2 D_{ij} dt
\end{align}
obtained from the LE to the same expectation derived from the change of variable $\pmb{x}_{dt}=\pmb{y}_{dt}+\pmb{v}\, dt$,
\begin{align}\label{proof_B2b}
\langle x^{i}_{dt}  x^{j}_{dt} \rangle 
&= \langle y^{i}  y^{j} \rangle + 2 \left[v^{i} \langle y^{j} \rangle\right]\sym dt + O(dt^2)\,,
\end{align}
Since $\langle x^{i}_{0} ~\! x^{j}_{0} \rangle=\langle y^{i} ~\! y^{j} \rangle$, from \eqref{proof_B2b} and \eqref{proof_B2a} to order $dt$,
one gets
\begin{equation}\label{proof_B2}
\left[v^{i} \langle y^{j} \rangle\right]\sym = \left[\mu_{jk}~\!\langle x^{i}_{0} ~\! F^{k}_{0} \rangle\right]\sym + \! D_{ij}~\! \, .   
\end{equation}
which has the same right-hand side of \eqref{proof_B1}. Hence, from \eqref{proof_B1} and \eqref{proof_B2}, we finally find
\begin{equation}\label{xnu_0}
    \left[\langle x^{i}_{0} ~ \nu^{j}_{0} \rangle\right]\sym 
    = \left[v^{i} \langle y^{j} \rangle\right]\sym 
    = \left[\mean{x_0^i} \mean{\nu_0^j} \right]\sym \, .
\end{equation}
showing that the symmetrized covariance of $x_0^i$ and $\nu_0^j$ is zero.

\section{Excess variance as a function of the mean local velocity}
\label{app:Deriv_vel}

We show that the components $\EV_{ij}(t)$ of the excess (co)variance can be rewritten as a double-time integral of the symmetrized connected correlation function
\begin{equation}
    \left[C^{ij}_{\dot x \nu }(t) \right]\sym = \frac{C^{ij}_{\dot x \nu }(t) + C^{ji}_{\dot x \nu }(t)}{2}
    =
    \frac 1 2
    \left(
    \mean{\dot x_t^i \nu_0^j } 
    -\mean{\dot x_t^i}\mean{\nu_0^j }
    + \mean{\dot x_t^j \nu_0^i } 
    -\mean{\dot x_t^j}\mean{\nu_0^i }
    \right)
\end{equation}
of actual velocity $\pmb{\dot x}_t$ and mean local velocity $\pmb{\nu}_0 = \pmb{\nu}(\pmb{x}_0)$:
\begin{align}
\label{Sij1}
   \EV_{ij}(t) & =
   2 \int_{0}^{t}\mathrm{d}t'\left[\mathrm{Cov} \left( x^{i}_{t}-x^{i}_{0} ~\!,~\!\mu_{jk}\, F^{k}_{t'} \right)\right]\sym 
   \\&= 
4\int_{0}^{t}\mathrm{d}t'\!\int_{0}^{t'}\mathrm{d}t'' \left[C^{ij}_{\dot x \nu }(t'') \right]\sym \,.
\label{Sij2}
\end{align}
To rewrite the integral in \eqref{Sij1} as twice the double integral in \eqref{Sij2}, we start by noting from the LE \eqref{eq:LE} that
\begin{align}
 \mu_{ij} \int_{0}^{t} \mathrm{d}t' ~ F^{j}_{t'} = x^{i}_t - x^{i}_0 - \sqrt{2~\!D_{ij}}\int_{0}^{t} \mathrm{d}t'~ \xi^{j}_{t'} \, , 
\end{align}
leads to
\begin{align}
\label{eq:S-step1}
\frac 1 2 \EV_{ij}(t) 
& =
\left[\mu_{jk}\! \int_{0}^{t}\mathrm{d}t'~ \mathrm{Cov} \left( x^{i}_{t}-x^{i}_{0} ~\!,~\! F^{k}_{t'} \right)\right]\sym
\\
\label{eq:S-step2}
&=
~C^{ij}_{\Delta x}(t)-\left[\sqrt{2~\!D_{jk}}\int_{0}^{t}\mathrm{d}t'~\! \langle x^{i}_{t} ~ \xi^{k}_{t'} \rangle \right]\sym
\\
\label{eq:S-step3}
&=
~C^{ij}_{\Delta x}(t)  - \left[\sqrt{ 2~\! D_{jk}\!} \int_{0}^{t}\mathrm{d}t'~\! \bigg \langle \frac{\delta ~\! x^{i}_{t} }{\delta\xi^{k}_{t'}} \bigg \rangle \right]\sym 
\\
\label{eq:S-step4}
& = 
~C^{ij}_{\Delta x}(t) - 2\left[
\langle x^{i}_{t}~ x^{j}_{t}\rangle
-\langle x^{i}_{t}~ x^{j}_{0}\rangle-\int_{0}^{t}\mathrm{d}t'~\!\langle x^{i}_{t} ~ \nu^{j}_{t'} \rangle \right]\sym 
\\
\label{eq:S-step5}
& = \langle x_{0}^{i}~x_{0}^{j} \rangle
-\langle x_{t}^{i}~x_{t}^{j} \rangle
+2 \int_{0}^{t}\mathrm{d}t'\left[\langle x^{i}_{t} ~ \nu^{j}_{t'} \rangle\right]\sym - \langle x^{i}_{t}-x^{i}_{0} \rangle \langle x^{j}_{t}-x^{j}_{0} \rangle \, .
\end{align}
To get from \eqref{eq:S-step1} to \eqref{eq:S-step2}, we use $\mathrm{Cov} ( x^{j}_{t} ~\!,~\!  \xi^{i}_{t'} ) = \langle x^{j}_{t} ~\!  \xi^{i}_{t'} \rangle$ (because $\langle \xi^{i}_{t'} \rangle = 0$) and $\int_{0}^{t}\mathrm{d}t'~\! \langle x^{i}_{0} ~ \xi^{j}_{t'} \rangle =0$ since $\langle x^{i}_{0} ~ \xi^{j}_{t'} \rangle=0$ almost everywhere on $[ 0,t]$ for every $\{ij\}$. The step to \eqref{eq:S-step3} derives from the Furutsu-Novikov formula \cite{Furutsu,Novikov}, which relates the position-noise correlation $\langle x^{i}_{t} ~ \xi^{k}_{t'} \rangle$ to the response function \eqref{resp_func} derived in \ref{app:R}.
In~\eqref{eq:S-step4} we replace the response function~\eqref{Rik}, while the last step to~\eqref{eq:S-step5} uses the expansion 
\begin{equation}
     \mathrm{Cov}( x^{i}_{t}-x^{i}_{0}~,~x^{j}_{t}-x^{j}_{0} ) =  \langle x_{t}^{i}~x_{t}^{j} \rangle+\langle x_{0}^{i}~x_{0}^{j} \rangle  -\langle x^{j}_{t}~ x^{i}_{0}\rangle-\langle x^{i}_{t}~ x^{j}_{0}\rangle-\langle x^{i}_{t}-x^{i}_{0} \rangle \langle x^{j}_{t}-x^{j}_{0} \rangle \,.
\end{equation}

When expressing correlations as a function of the position $\pmb{y}_t = \pmb{x}_t - \pmb{v}t$ (see~\ref{app:y}), several terms drop out from \eqref{eq:S-step5},
\begin{align}
\label{eq:S-step6}
    \frac 1 2 \EV_{ij}(t) 
    &=
2\int_{0}^{t}\mathrm{d}t'\left[\mean{y^i_t \nu^j_{t'}}\right]\sym
   -2 t \left[\mean{y^i_t} v^j\right]\sym 
   \nonumber\\
    &=
2\int_{0}^{t}\mathrm{d}t'\left[\mean{ y^i_t \nu^j_{t'}} -\mean{y^i_t}\mean{\nu^j_{t'}} \right]\sym
   \nonumber\\
    &=
2\int_{0}^{t}\mathrm{d}t'\left[\mean{ y^i_{t'} \nu^j_0} -\mean{y^i_{t'}}\mean{\nu^j_0} \right]\sym
   \nonumber\\
    &=
   2 \int_{0}^{t}\mathrm{d}t'\left[ C^{ij}_{y \nu }(t')\right]\sym
   \nonumber\\
    &=
2\int_{0}^{t}\mathrm{d}t'\left[ C^{ij}_{x \nu }(t')\right]\sym\,.
   \nonumber\\
    &=
2\int_{0}^{t}\mathrm{d}t'\int_{0}^{t'}\mathrm{d}t''\left[ C^{ij}_{\dot x \nu }(t'')\right]\sym\,.
\end{align}
The steps above conclude the proof by using the process' stationarity and the fact that variables' covariance does not change when they are displaced by a constant ($\pmb{v}t'$ bringing $\pmb{y}_{t'}$ to $\pmb{x}_{t'}$). The last equality uses \eqref{zero_cov}
.

\section{VSR of the Brownian gyrator}
\label{app:BG}

We show how to compute the terms of the VSR for the model of Brownian gyrator.
We apply the Laplace transform to its equation of motion \eqref{LE_brow_gyr}. By turning to matrix notation and by doing some algebra, we get
\begin{equation}\label{Brow_gyr_sol_step1}
\begin{pmatrix}
\hat{x}_s\\[5pt]
\hat{y}_s
\end{pmatrix}= \hat{\pmb{\chi}}_s \cdot
\begin{pmatrix}
x_{0}+\sqrt{ 2\,k_{B}T_{1}\,\mu_{1}}~\hat{\xi}^{x}_s\\[5pt]
y_{0}+\sqrt{ 2\,k_{B}T_{2}\,\mu_{2}}~\hat{\xi}^{y}_s
\end{pmatrix}\, ,
\end{equation}
where this time, the susceptibility matrix is equal to
\begin{equation}
   \hat{\pmb{\chi}}_s =  \frac{1}{(s+\mu_{1}\,\kappa)(s+\mu_{2}\,\kappa)- \alpha^2\,\mu_{1}\,\mu_{2}\,\kappa^{2}} 
\begin{pmatrix}
s+\mu_{2}\,\kappa & -\alpha \,\mu_{1} \, \kappa \\[5pt]
-\alpha \,\mu_{2} \, \kappa & s + \mu_{1} \, \kappa  
\end{pmatrix} \, .
\end{equation}
We define the function
\begin{equation}
\begin{split}
    \mathcal{T}_t =& \mathcal{L}^{-1}\!\left[\frac{1}{(s+\mu_{1}\,\kappa)(s+\mu_{2}\,\kappa)- \alpha^2\,\mu_{1}\,\mu_{2}\,\kappa^{2}}  \right] = \\[7pt]
    =&\frac{2 \,  \sinh \left(\kappa \, t  \sqrt{4 \alpha^2 \mu_{1} \,\mu_{2}+(\mu_{1}-\mu_{2})^2}/2\right)}{\kappa  \sqrt{4 \alpha^2 \mu_{1} \,\mu_{2}+(\mu_{1}-\mu_{2})^2}}\text{e}^{-t \, \kappa \left( \mu_{1} + \mu_{2}\right)/2} \, ,
\end{split}
\end{equation}
which is such that $\mathcal{T}(0) = 0$ (implying that $\overset{\,\bm.}{\mathcal{T}}\,{\vphantom{\mathcal{T}}}_t = \mathcal{L}^{-1}\left[s\,\hat{\mathcal{T}}_s \right]$). It can be used to express the susceptibility matrix in real time as
\begin{equation}\label{chi_BG}
\pmb{\chi}_t =\begin{pmatrix}
\overset{\,\bm.}{\mathcal{T}}\,{\vphantom{\mathcal{T}}}_t+\mu_{2}\,\kappa\,\mathcal{T}_t & -\alpha \,\mu_{1} \, \kappa\,\mathcal{T}_t \\[5pt]
-\alpha \,\mu_{2} \, \kappa \,\mathcal{T}_t & \overset{\,\bm.}{\mathcal{T}}\,{\vphantom{\mathcal{T}}}_t + \mu_{1} \, \kappa  \,\mathcal{T}_t 
\end{pmatrix} \, .
\end{equation}
Furthermore, by taking the inverse Laplace transform of \eqref{Brow_gyr_sol_step1}, the solution of \eqref{LE_brow_gyr} can be written as
\begin{equation}\label{sol_LE_Brow_gyr}
    x^{i}_t = \chi^{ij}_t \, x^{j}_{0} + \int_{0}^{t}\mathrm{d}t' \, \chi^{ij}_{t-t'} \, \xi^{j}_{t'} \, .
\end{equation}
From this, one can readily evaluate the stationary average of $\pmb{x}_{t}$
\begin{equation}\label{avg_Br_gyr}
    \langle \pmb{x} \rangle^{\text{st}}_{t} = \langle \pmb{x} \rangle^{\text{st}}_{0} = \pmb{\chi}_t\, \langle \pmb{x} \rangle_{0}^{\text{st}}\, ,
\end{equation}
where $\langle \pmb{x} \rangle^{\text{st}}_{t} = \langle \pmb{x} \rangle^{\text{st}}_{0}$ is a consequence of the fact that neither the drift vector nor the diffusion matrix explicitly depends on time. Moreover, because the matrix $\pmb{\chi}_t\neq\mathbb{1}_{2}$ is not degenerate, the only possibility for \eqref{avg_Br_gyr} to hold corresponds to $\langle \pmb{x} \rangle_{0}^{\text{st}} = 0$. Another consequence of \eqref{sol_LE_Brow_gyr} is that if $\pmb{x}_{0}$ is distributed as a bivariate Gaussian, the same will be true for $\pmb{x}_{t}$ at all times because a sum (or integral) of Gaussian random variables is itself Gaussian (remember that $\pmb{\xi}_t$ is also Gaussian). This feature is indeed a consequence of the linearity of equation \eqref{LE_brow_gyr} and, along with $\langle \pmb{x} \rangle_{0}^{\text{st}} = 0$, it implies that 
\begin{equation}\label{p_st_brow_gyr}
    p^{\text{st}}_t \sim \exp \left( -\frac{1}{2} \pmb{x}^{\text{T}}_{t}~ {\cal V}_{\pmb{x}}^{-1} ~ \pmb{x}_{t}  \right) \, ,
\end{equation}
with covariance matrix
\begin{equation}\label{cov_pos_0_br_gyr}
    {\cal V}_{\pmb{x}} = 
    \begin{pmatrix}
    \langle  x_{0}^{2} \rangle & \langle x_{0}\,y_{0} \rangle\\[5pt]
    \langle x_{0}\,y_{0} \rangle & \langle y_{0}^{2} \rangle
    \end{pmatrix} \,.
\end{equation}
In order to evaluate its components, we resort to the discrete-time version of \eqref{LE_brow_gyr},
\begin{equation}\label{LE_brow_gyr_disc}
\begin{cases}
x_{t+dt} = x_t -\mu_1\,\kappa\,x_t\,dt -\mu_1\,\alpha\,\kappa\,y_t\,dt + \sqrt{ 2\,k_{B}T_{1}\,\mu_{1}} \,d\mathcal{B}^{x}_t\hspace{0.5cm}(a)\\
y_{t+dt} = y_t -\mu_2\,\kappa\,y_t\,dt -\mu_2\,\alpha\,\kappa\,x_t\,dt + \sqrt{ 2\,k_{B}T_{2}\,\mu_{2}} \,d\mathcal{B}^{y}_t\hspace{0.5cm}(b)\\
\end{cases}
\end{equation}
where $\langle d\mathcal{B}^{i}_t \rangle = 0$ and $\langle d\mathcal{B}^{i}_t \, d\mathcal{B}^{j}_t \rangle = \delta_{ij} ~ dt$. By taking the square of (a) and (b) along with the product between (a) and (b) and by taking their average, one gets
\begin{equation}
\begin{cases}
\langle x^2_{t+dt} \rangle = \langle x^2_{t} \rangle + \left( -2\,\mu_1\,\kappa\,\langle x^{2}_{t} \rangle -2\,\mu_1\,\alpha\,\kappa\,\langle y_{t} \, x_{t} \rangle +  2\,k_{B}T_{1}\,\mu_{1} \right)\,dt + o(dt)\\
\langle y^2_{t+dt} \rangle = \langle y^2_{t} \rangle + \left( -2\,\mu_2\,\kappa\,\langle y^{2}_{t} \rangle -2\,\mu_2\,\alpha\,\kappa\,\langle y_{t} \, x_{t} \rangle +  2\,k_{B}T_{2}\,\mu_{2} \right)\,dt + o(dt)\\
\langle x_{t+dt} \, y_{t+dt} \rangle = \langle x_{t} \, y_{t} \rangle - (\kappa (\mu_{1}+\mu_{2}) \langle x_{t}\, y_{t} \rangle + \alpha \, \mu_{2} \, \kappa \langle x^{2}_{t} \rangle+ \alpha \, \mu_{1} \, \kappa \langle y^{2}_{t} \rangle)dt+o(dt) \, .
\end{cases}
\end{equation}

By further noting that, in a steady state, the correlation functions of random variables with constant average only depend on time differences (meaning, for example that $\langle x^2_{t+dt} \rangle = \langle x^2_{t} \rangle$), one readily sees that \eqref{LE_brow_gyr_disc} leads to a linear system of three equations and three variables (i.e. $\langle x^{2}_{t} \rangle$, $\langle y^{2}_{t} \rangle$ and $\langle x_{t}\,y_{t} \rangle$, which are of course constant in time) whose solution reads
\begin{equation}\label{br_gyr_init_cond}
\begin{split}
\langle x^{2}_{0} \rangle =&~ k_{B}\frac{T_{1}(\mu_1+\mu_2)+\alpha^{2}\,\mu_{1}\,(T_{2}-T_{1})}{\kappa(1-\alpha^{2})(\mu_1+\mu_2)}\\[4pt]
\langle y^{2}_{0} \rangle =&~ k_{B}\frac{T_{2}(\mu_1+\mu_2)+\alpha^{2}\,\mu_{2}\,(T_{1}-T_{2})}{\kappa(1-\alpha^{2})(\mu_1+\mu_2)}\\[4pt]
\langle x_{0}\,y_{0} \rangle =& - \alpha\, k_{B}\frac{\mu_{1}\,T_{2}+ \mu_{2}\,T_{1}}{\kappa(1-\alpha^{2})(\mu_1+\mu_2)}\, .
\end{split}
\end{equation}
These quantities  fully determine the stationary PDF $p^{\text{st}}_t$ \eqref{p_st_brow_gyr}, which in turn allows us to calculate the mean local velocity
\begin{equation}\label{mean_loc_vel_Br_gyr}
    \pmb{\nu}_{t} = -\pmb{\mu}\, \nabla U_t - \pmb{D} \, \nabla \ln p^{\text{st}}_{t} = \begin{pmatrix}
    a_{1}\,x_{t}-b_{1}\,y_{t} \\[5pt]
    a_{2}\,y_{t}-b_{2}\,x_{t}
    \end{pmatrix} \, ,
\end{equation}
where $(\pmb{\mu})_{ij} = \mu_{i}\delta_{ij}$ is the mobility matrix and
\begin{equation}\label{mean_loc_vel_param_br_gyr}
\begin{split}
    a_{1} = \frac{k_{B}T_{1}\,\mu_{1}\langle y_{0}^{2} \rangle}{\langle x_{0}^{2} \rangle\langle y_{0}^{2} \rangle-\langle x_{0}\,y_{0} \rangle^{2}}-\mu_{1}\,\kappa \,, \hspace{1cm}& b_{1} = \frac{k_{B}T_{1}\,\mu_{1}\langle x_{0}\,y_{0} \rangle}{\langle x_{0}^{2} \rangle\langle y_{0}^{2} \rangle-\langle x_{0}\,y_{0} \rangle^{2}}-\alpha\,\mu_{1}\,\kappa\,, \\[5pt]
     a_{2} = \frac{k_{B}T_{2}\,\mu_{2}\langle x_{0}^{2} \rangle}{\langle x_{0}^{2} \rangle\langle y_{0}^{2} \rangle-\langle x_{0}\,y_{0} \rangle^{2}}-\mu_{2}\,\kappa \,, \hspace{1cm}& b_{2} = \frac{k_{B}T_{2}\,\mu_{2}\langle x_{0}\,y_{0} \rangle}{\langle x_{0}^{2} \rangle\langle y_{0}^{2} \rangle-\langle x_{0}\,y_{0} \rangle^{2}}-\alpha\,\mu_{2}\,\kappa \, .
\end{split}
\end{equation}
With this, it is straightforward to obtain $\sigma$, which reads
\begin{equation}\label{entr_Br_gyr}
    \sigma = 
    \big\langle \pmb{\nu}_t ~ \pmb{D}^{\text{-}1} \, \pmb{\nu}_t \big \rangle 
    = \alpha^{2}\,\kappa\,\frac{\mu_{1}\,\mu_{2}}{
    \mu_{1}+\mu_{2} }\frac{(T_{1}-T_{2})^{2}}{
T_{1}\,T_{2}}  \, 
\end{equation}
and equals \eqref{sigma_BG}.

The next step consists in the calculation of the position correlation functions  
\begin{equation}\label{corr_pos_br_gyr}
\begin{split}
    \langle \pmb{x}_{t} \, \pmb{x}^{\text{T}}_{0}\rangle  = & \pmb{\chi}_t \langle  \pmb{x}_{0} \,  \pmb{x}^{\text{T}}_{0}\rangle = \pmb{\chi}_t {\cal V}_{\pmb{x}} \\[7pt]
    =& 
    \begin{pmatrix}
 \begin{matrix}
 \left( \overset{\,\bm.}{\mathcal{T}}\,{\vphantom{\mathcal{T}}}_t+ \kappa\,\mu_{2}\,\mathcal{T}_t \right)\langle x^{2}_{0} \rangle+{}\\[2pt] -\alpha\,\kappa\,\mu_{1}\,\mathcal{T}_t\langle x_{0}\, y_{0} \rangle
 \end{matrix} & \hspace{1cm}
 \begin{matrix}
 \left( \overset{\,\bm.}{\mathcal{T}}\,{\vphantom{\mathcal{T}}}_t+ \kappa\,\mu_{2}\,\mathcal{T}_t \right)\langle x_{0}\, y_{0} \rangle+{}\\[2pt] -\alpha\,\kappa\,\mu_{1}\,\mathcal{T}_t\langle y^{2}_{0} \rangle
 \end{matrix}
  \\[30pt]
  \begin{matrix}
  \left( \overset{\,\bm.}{\mathcal{T}}\,{\vphantom{\mathcal{T}}}_t+ \kappa\,\mu_{1}\,\mathcal{T}_t \right)\langle x_{0}\, y_{0} \rangle+{}\\[2pt] -\alpha\,\kappa\,\mu_{2}\,\mathcal{T}_t\langle x^{2}_{0} \rangle
  \end{matrix}
  & \hspace{1cm}
  \begin{matrix}
  \left( \overset{\,\bm.}{\mathcal{T}}\,{\vphantom{\mathcal{T}}}_t+ \kappa\,\mu_{1}\,\mathcal{T}_t \right)\langle y^{2}_{0} \rangle+{}\\[2pt] -\alpha\,\kappa\,\mu_{2}\,\mathcal{T}_t\langle x_{0}\, y_{0} \rangle
  \end{matrix}
\end{pmatrix}\, ,
\end{split}
\end{equation}
where in the first equality we used \eqref{sol_LE_Brow_gyr} along with the fact that $\langle \pmb{\xi}(t')~\pmb{x}_{0} \rangle=0$ almost everywhere and then we combined \eqref{chi_BG} and\eqref{cov_pos_0_br_gyr}. Note that, because $\mathcal{T}(0)=0$ and $\overset{\,\bm.}{\mathcal{T}}\,{\vphantom{\mathcal{T}}}(0)=1$, for $t=0$ \eqref{corr_pos_br_gyr} trivially reduces to \eqref{cov_pos_0_br_gyr}. Again, because the system is linear, these correlations turn out to be the building blocks for the computation of all the quantities appearing in the VSR. In particular, for the variance of the relative displacement, we find that
\begin{equation}
\begin{split}
{\cal V}_{\Delta \pmb{x}}(t)& 
= 2 \langle \pmb{x}_{0} \, \pmb{x}^{\text{T}}_{0}\rangle-\langle \pmb{x}_{t} \, \pmb{x}^{\text{T}}_{0}\rangle-\langle \pmb{x}_{t} \, \pmb{x}^{\text{T}}_{0}\rangle^{\mathrm{T}}  \\[7pt]
=&2 \begin{pmatrix}
\langle x^{2}_{0} \rangle -\langle x_{t}\, x_{0} \rangle & \langle x_{0}\,y_{0} \rangle -\left(\langle x_{t}\,y_{0} \rangle + \langle y_{t}\,x_{0} \rangle  \right)/2 \\[5pt]
\langle x_{0}\,y_{0} \rangle -\left(\langle x_{t}\,y_{0} \rangle + \langle y_{t}\,x_{0} \rangle  \right)/2 & \langle y^{2}_{0} \rangle -\langle y_{t}\, y_{0} \rangle
\end{pmatrix} \, ,
\end{split}
\end{equation}
whereas for the covariance matrix of the integrated forces, with $F_t = -\nabla U_t$, we obtain

\begin{equation}\label{cov_sum_forc_br_gyr}
\begin{split}
&{\cal V}_{\Sigma_{\pmb{F}}}(t) =\int_{0}^{t}\mathrm{d}t'\int_{0}^{t}\mathrm{d}t''  \big \langle \pmb{F}_{t'} \, \pmb{F}^{\mathrm{T}}_{t''} \big\rangle\\[8pt]
&\langle \pmb{F}_{t'} \, \pmb{F}^{\mathrm{T}}_{t''} \big\rangle= 2\,\kappa^{2}\begin{pmatrix}
\begin{matrix}
      \mu_{1}^{2}\big( \langle x_{t''}\,x_{0}\rangle + \alpha^{2} \langle y_{t''}\,y_{0}\rangle+ {} \\[2pt]
      +\alpha (\langle x_{t''}\,y_{0} \rangle+\langle y_{t''}\,x_{0} \rangle) \big)
    \end{matrix}
 & 
 \begin{matrix}
      \mu_{1}\,\mu_{2}\big( \alpha(\langle x_{t''}\,x_{0}\rangle +  \langle y_{t''}\,y_{0}\rangle)+ {} \\[2pt]
      +(1+\alpha^{2}) (\langle x_{t''}\,y_{0} \rangle+\langle y_{t''}\,x_{0} \rangle)/2 \big)
    \end{matrix}
 \\[30pt]
\begin{matrix}
      \mu_{1}\,\mu_{2}\big( \alpha(\langle x_{t''}\,x_{0}\rangle + \langle y_{t''}\,y_{0}\rangle)+ {} \\[2pt]
      +(1+\alpha^{2}) (\langle x_{t''}\,y_{0} \rangle+\langle y_{t''}\,x_{0} \rangle)/2 \big)
    \end{matrix}
 & 
 \begin{matrix}
      \mu_{2}^{2}\big( \langle y_{t''}\,y_{0}\rangle + \alpha^{2} \langle x_{t''}\,x_{0}\rangle+ {} \\[2pt]
      +\alpha (\langle x_{t''}\,y_{0} \rangle+\langle y_{t''}\,x_{0} \rangle) \big)
    \end{matrix}
\end{pmatrix} \, .
\end{split}
\end{equation}
Finally, for the excess variance, we can use \eqref{eq:Sv} for example to obtain
\begin{equation}\label{viol_fact_br_gyr}
\begin{split}
    \pmb{\EV}(t) =4\bigintssss_{0}^{t}\mathrm{d}t'
\begin{pmatrix}
a_{1}\,\langle x_{t'}\,x_{0}\rangle-b_{1}\langle x_{t'} \,y_{0} \rangle & \hspace{0.5cm}
\begin{matrix}
     \Big(a_{1}\, \langle y_{t'} x_{0} \rangle + a_{2}\, \langle x_{t'} y_{0} \rangle +{}\\[2pt]
     -b_{1}\langle y_{t'} y_{0} \rangle -b_{2}\langle x_{t'} x_{0} \rangle \big)/2
\end{matrix}
\\[20pt]
\begin{matrix}
     \Big(a_{1}\, \langle y_{t'} x_{0} \rangle + a_{2}\, \langle x_{t'} y_{0} \rangle +{}\\[2pt]
     -b_{1}\langle y_{t'} y_{0} \rangle -b_{2}\langle x_{t'} x_{0} \rangle \big)/2
\end{matrix} & \hspace{0.5cm}
a_{2}\,\langle y_{t'}\,y_{0}\rangle-b_{2}\langle y_{t'} \,x_{0} \rangle
\end{pmatrix}
\end{split}
\end{equation}
where $a_1,\, a_2,\, b_1,\, b_2$ are given by \eqref{mean_loc_vel_param_br_gyr}.

\section*{Bibliography}
\providecommand{\newblock}{}

\end{document}